\documentclass{emulateapj}

\def\kms{km~s$^{-1}$}
\def\ga{\mathrel{\hbox{\rlap{\hbox{\lower4pt\hbox{$\sim$}}}\hbox{$>$}}}}
\def\la{\mathrel{\hbox{\rlap{\hbox{\lower4pt\hbox{$\sim$}}}\hbox{$<$}}}}

\shorttitle{Turbulence properties of HI in the SMC}

\shortauthors{BURKHART ET AL.}

\begin{document}

\title{Characterizing Magnetohydrodynamic Turbulence in the Small Magellanic Cloud}

\author{Blakesley Burkhart\altaffilmark{1}, Sne\v{z}ana Stanimirovi\'c\altaffilmark{1}, 
Alex Lazarian\altaffilmark{1}, \& Grzegorz Kowal\altaffilmark{1,2}}
\altaffiltext{1}{Astronomy Department, University of Wisconsin, Madison, 475 N. 
Charter St., WI 53711, USA}
\altaffiltext{2}{Astronomical Observatory, Jagiellonian University, 
ul. Orla 171, 30-244 Krak\'ow, Poland}

\begin{abstract} 
We investigate the nature and spatial variations of turbulence in the 
Small Magellanic Cloud (SMC) by applying several statistical methods 
on  the neutral hydrogen (HI) 
column density image of the SMC and a database of isothermal numerical simulations.
By using the 3rd and 4th statistical moments we derive the spatial distribution of
the sonic Mach number (${\cal M}_s$) across the SMC. We find that about 90\% of the HI
in the SMC is subsonic or transonic. However, edges of the SMC `bar' have
${\cal M}_s\sim4$ and may be tracing shearing or turbulent flows.
Using numerical simulations we also investigate how the slope of the spatial power spectrum
depends on both sonic and Alfv\'en Mach numbers. 
This allows us to gauge the Alfv\'en Mach number of the SMC and conclude
that its gas pressure dominates over the magnetic pressure.
The super-Alfv\'enic nature of the HI gas in the SMC is also highlighted
by the bispectrum, a three-point correlation function
which characterizes the level of non-Gaussianity in wave modes.
We find that the bispectrum of the SMC HI column density
displays similar large-scale  correlations as numerical simulations,
however it has localized enhancements of correlations. In addition, we find a break
in correlations at a scale of $\sim160$ pc. This may be caused by
numerous expanding shells of a similar size.
\end{abstract}

\section{Introduction}

\label{intro}

In the recent decade, many advances in both observations and computational models 
have provided new insights into the workings and evolution of the interstellar medium (ISM).
The emerging picture is that interstellar turbulence 
plays the key role in ISM structure formation and evolution 
(McKee \& Ostriker 2007). In the Galaxy and the Magellanic Clouds, 
the ISM is turbulent on scales ranging from less than a parsec
to a few kiloparsecs (Crovisier \& Dickey 1983; 
Stanimirovic et al. 1999; Deshpande et al. 2000; Dickey et al. 2001; 
Elmegreen et al. 2001; Elmegreen \& Scalo 2004). 
Although the observational evidence for the importance of turbulence in 
the ISM is overwhelming, many questions
remain open. For example, what are the dominant energy sources and physical processes that convert
kinetic energy into turbulence (Burkert 2006)? At what scales and through which modes is turbulent
energy dissipated (Heyer \& Zweibel 2004)? 
How do the level and type of turbulence depend on properties
of the interstellar gas (e.g. presence/absence of star formation, 
presence/absence of tidal effects, or the strength of magnetic field)?
Since no complete theory of astrophysical turbulence exists, studying its 
effects on the multiphase ISM can be challenging and calls for a combination of 
numerical and observational efforts.

Statistical studies have proved to be important in characterizing  
the magnetized turbulent ISM  \cite[]{Lazarian09}, however the interpretation of results
is not always straight forward. 
Several statistical methods have been extensively used for both observational and synthetic data.  
These statistics include probability density functions (PDFs), wavelets, the principal
component analysis,  higher order moments, 
Velocity Coordinate Spectrum (VCS), and Velocity Channel 
Analysis (VCA), to name just a few (Gill \& Henriksen
1990; Brunt \& Heyer 2002; Kowal, Lazarian
\& Pogosyan 2000, 2004, 2006;  Lazarian \& Beresnyak 2007).

Most of these statistical methods require large datasets with a large 
spatial or velocity dynamic range, 
and produce a single, mostly one-dimensional, measure. 
This results in the lack of spatial information
about turbulent properties across a given interstellar cloud, or a galaxy, 
making a connection with underlying physical properties highly difficult.

In this paper we explore a new method for obtaining spatial information about the level 
and nature of ISM
turbulence on the neutral hydrogen (HI) observations of the Small Magellanic Cloud (SMC).
The SMC, a dwarf irregular galaxy in the Local Group, has 
a highly gas rich ISM environment (see, Stanimirovic et al. 1999, 
henceforth known as SX99), and is an excellent candidate for ISM studies.
Being nearby (60 kpc, \cite{West91}), the SMC is distant enough for 
all its objects to be treated as having roughly the same 
distance, unlike the Milky Way where distance determination is relatively uncertain.

The HI observations of the SMC, obtained using the Australia Telescope Compact Array (ATCA) and 
the Parkes telescope (SX99),
have been used for several investigations,  including the HI spatial power 
spectrum and the kinematic study of HI, which revealed the existence of many expanding 
shells of gas and three supergiant shells. 
The power law index of HI density and velocity distributions was derived in SX99 and \cite{Stan01}, 
while the Genus statistic in \cite{Chep08} revealed spatial variations of HI morphology. 
Because the SX99 SMC data set is  well studied, it is a perfect candidate to 
investigate new statistical methods. 
We can acquire new information, but also test and confirm past results, 
as well as validate the promise of these statistical tools for 
further use in other observational studies.

In this study, we investigate turbulent properties of the HI in the SMC by applying 
the higher order moments on the HI column density image.
We then use a database of MHD simulations to bootstrap
the spatial distribution of the sonic Mach number across the SMC.
The crucial aspect of our approach is the confluence of observations
and numerical simulations: only by combining the two
we can retrieve the spatial variations of turbulent properties. 
This is the reason why we oscillate between observational and synthetic data
in this paper.
We also investigate whether and how interstellar shocks
leave footprints on the HI gas by employing the bispectrum,
a three point statistical measure, on the SMC HI column density image.
Again, to interpret our results we apply the same statistics on
the database of MHD simulated column density images.

In particular, the paper is organized as follows. 
We start with \S~\ref{sec:background} by providing a brief summary of previous work
regarding the statistical methods used in our study.
In \S~\ref{sec:data} we describe the SMC HI column density map and the 
database of numerical 
simulations of compressible MHD turbulence used for the comparison with observations. 
In \S~4 we introduce higher order moments and their dependence on the sonic
and Alfvenic Mach numbers. We then apply higher order moments on the
SMC HI observations to derive an image of the sonic Mach number across
the SMC, in \S~5. 
We  compare our results with an observational estimate of 
the sonic Mach number of the cold neutral medium (CNM) in the SMC, based
on a comparison of the spin and kinetic temperature of HI absorption profiles,
 in \S~\ref{ratio}. 
In \S~\ref{ps} we show how the power-law slope of the spatial power spectrum  
depends on the sonic Mach number and use this to gauge the Alfvenic Mach number of the 
HI gas seen in emission. 
In \S~\ref{sec:bispectrum} we present an analysis of the bispectrum of the SMC, 
as well as a brief discussion of the noise and windowing effects. 
In \S~\ref{sec:Discussion} we provide a discussion of our results, followed by 
our conclusions.

\section{Background on statistical methods used in this study}
\label{sec:background}

\subsection{Higher order statistical moments}

Higher order statistical moments of density fluctuations 
have been studied extensively.
For example, the variance of density fluctuations has been shown to increase with 
the sonic Mach number ${\cal M}_s$ \cite[]{Nor99,Ostriker01}.
Therefore, if turbulence is the dominant structuring mechanism, ${\cal M}_s$ can be estimated 
from the variance of density fluctuations.
However, the observable that is the most easily available from observations
for various ISM tracers is the column density. While this is 
a less direct measure of turbulence compared to velocity,
a comparison between observations and simulations is the most straight forward
in case of column densities.

Only very recently, Kowal et al. (2007), henceforth referred to as KLB, 
have investigated how variance, skewness and kurtosis, the 2nd, 3rd and 4th 
order moments respectively, depend on ${\cal M}_s$. They found strong
correlations: as the sonic Mach number increases, so does the Gaussian 
asymmetry of the column density (and density) PDFs due to gas 
compression via shocks, resulting in the increase of variance, skewness and kurtosis. 
KLB used limited resolution models of 128$^3$ in their study, while
Burkhart et al. (2009), henceforth known as BFKL, saw the same trends 
using high resolution isothermal simulations.  
In both studies, the moments had little dependence on the Alfv\'en Mach 
number (${\cal M}_A$), or the line-of-sight (LOS) orientation used for integrating
 3-D simulated data cubes.   
These studies are motivational, and imply that the sonic Mach number 
of turbulence in interstellar clouds could be characterized by variance, 
skewness and kurtosis of observed column density distributions.  

\subsection{Spatial power spectrum}

The two-dimensional spatial power spectrum characterizes the energy distribution over spatial 
scales.  SX99 found that the spatial power spectrum of individual HI velocity slices is well fit
by a power-law, with an average slope of $\approx -3$. 
Stanimirovic \& Lazarian (2001) showed that the power-law slope steepens
when several channels are integrated together, and used this to estimate 
the density power-law slope of $-3.3$ and the velocity slope of
$-3.4$. 
No obvious breaks or preferred scales were found over the range of 30 pc to 4 kpc. 
The density and velocity spectral slopes are similar, and in the case of incompressible MHD turbulence, density 
behaves as a passive scalar and thus scales in the same way as velocity (Monin \& Yaglom(1967), Lithwick \& Goldreich
(2001), Cho \& Lazarian 2003).
However, the estimated slopes
are slightly more shallow than the predictions for the Kolmogorov spectrum
($~k^{-11/3}$), which is expected for incompressible fluids with a weak magnetic field.

However, different types of turbulence are expected to have different spectral slopes. 

For example, in  incompressible fluids with a strong magnetic field,
the spectrum is expected to be even steeper and scale as  $~k^{-13/3}$ 
(Biskamp 2003). 

When the medium is supersonic (as we will see later applies to parts of the SMC), 
these relations are no longer valid due to shocks forming highly asymmetric density structures.
KLB demonstrated that the spectral slope of MHD turbulence becomes more shallow 
with increasing ${\cal M}_{s}$.  This can be understood as shocks in supersonic turbulence 
create more small scale structure in density \cite[]{beresnyak05}.  
This behavior was found to be weakly dependent of the Alfv\'en Mach number.

\subsection{Bispectrum}

While the spatial power spectrum has long been applied on both observations and
simulations,  it essentially uses 
only the amplitude of the Fourier transform of the initial signal, while the phase
information is totally ignored.
The  bispectrum, however, is a three point statistical measure which incorporates both 
the amplitude \textit{and phase} of the correlation of signals in 
Fourier space.  Because of this, 
it can be used to search for nonlinear wave-wave interactions and characterize 
how shocks affect turbulent properties of the ISM. 
The bispectrum  has been used in cosmology and gravitational wave studies 
to detect departures from Gaussianity \cite[]{Fry98,Scoccimarro00,Liguori06},
and for the characterization of wave-wave interactions 
in laboratory plasmas \cite[]{Intrator89,Tynan01}. 
The first application of the bispectrum on synthetic astrophysical MHD turbulence was in BFKL.

BFKL found  a general correlation between the bispectrum of
2D column density and 3D density maps for
simulated data cubes of 512$^3$ resolution. Also,
supersonic models showed a much greater degree of correlation between 
structures of different scales than subsonic models.  
While comparing models with the same sonic Mach number, models with a 
stronger magnetic field (i.e. sub-Alfv\'enic) showed enhanced correlations.  
These results are encouraging and suggest that the bispectrum can be also 
used to provide insight into the nature of the turbulence cascade.

\section{Data: SMC and Numerical Simulations}

\label{sec:data}

\subsection{HI Observations of the SMC}

The small-scale HI structure of the SMC was observed with
the ATCA, a radio interferometer,
in a mosaicking mode (Staveley-Smith et al.
1997). Observations of the same area were obtained also with
the 64-m Parkes telescope, in order to map the distribution of
large-scale features. Both sets of observations were then combined
(see SX99), resulting in the final HI data
cube with angular resolution of 98$''$, velocity resolution of 1.65 \kms, 
and 1-$\sigma$ brightness temperature sensitivity of 1.3 K, to
the continuous range of spatial scales between 30 pc and
4.4 kpc. The velocity range covered with these observations
is 90-215 \kms. For details about the ATCA and Parkes
observations, data processing, and data combination 
(short spacings correction) see Staveley-Smith et al. (1997) and SX99.

\begin{figure*}[tbh]
\centering
\includegraphics[scale=.5]{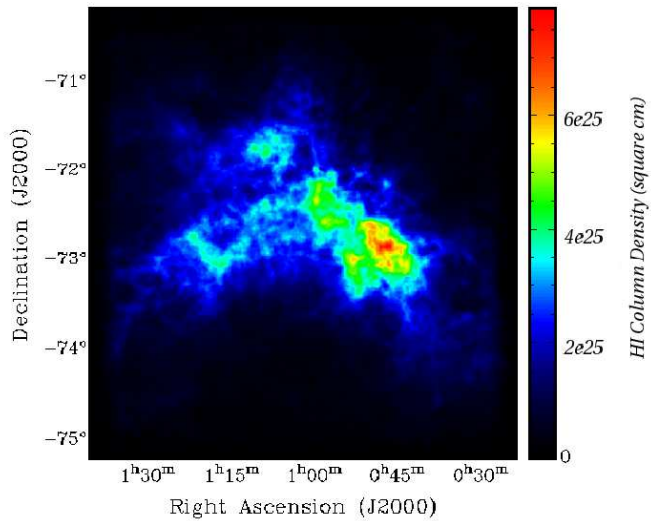}
\caption{The final HI column density image after correction for self-absorption. 
The largest scale is $\sim4.4$ kpc. The intensity range is in units 
of column density (cm$^{-2}$).}
\label{fig:smc}
\end{figure*}

The HI column density image is shown in Figure ~\ref{fig:smc}. 
We corrected the original image by
multiplying the HI column density of each pixel ($N_{HI}$ in atom cm$^{-2}$)
 with the correction factor $f_c$ derived by SX99:
\begin{displaymath}
 f_c = \left\{ \begin{array}{lr}
1+0.667(\log N_{HI}-21.4) & : \log N_{HI} > 21.4 \\
 1 & : \log N_{HI} \le 21.4
\end{array}
\right.
\end{displaymath} 

As the original Position-Position-Velocity (PPV) SMC cube has $578\times610\times78$ pixels, 
the resulting column density image is a 2D array with $578\times610$ pixels.

\subsection{A database of synthetic MHD cubes}

We generate a database of 13 3D numerical simulations of isothermal compressible (MHD)
turbulence by using the code of KLB and varying the input 
values for the sonic and Alfv\'enic Mach number.
The sonic Mach number is defined as ${\cal M}_s \equiv \langle |{\bf v}|/C_s \rangle$, 
where is ${\bf v}$ is the local velocity, $C_s$ is the sound speed, 
and the averaging is done over the whole simulation.
Similarly, the Alfv\'enic Mach number
is ${\cal M}_A\equiv \langle |{\bf v}|/v_A \rangle$, where 
$v_A = |{\bf B}|/\sqrt{\rho}$ is the Alfv\'enic velocity,
${\bf B}$ is magnetic field and $\rho$ is density.
KLB used resolution of 128$^3$, while we use 512$^3$.
Details about KLB's code were published (see Cho et al. 2003) and the code was used
in several studies. We briefly outline the major points of their numerical setup. 

The code is  a second-order-accurate hybrid essentially 
nonoscillatory (ENO) scheme \cite[see][]{Cho02} which solves
the ideal MHD equations in a periodic box:
\begin{eqnarray}
 \frac{\partial \rho}{\partial t} + \nabla \cdot (\rho {\bf v}) = 0, \\
 \frac{\partial \rho {\bf v}}{\partial t} + \nabla \cdot \left[ \rho {\bf v} {\bf v} + \left( p + \frac{B^2}{8 \pi} \right) {\bf I} - \frac{1}{4 \pi}{\bf B}{\bf B} \right] = {\bf f},  \\
 \frac{\partial {\bf B}}{\partial t} - \nabla \times ({\bf v} \times{\bf B}) = 0,
\end{eqnarray}
with zero-divergence condition $\nabla \cdot {\bf B} = 0$, 
and an isothermal equation of state $p = C_s^2 \rho$, where 
$p$ is the gas pressure. 
On the right-hand side, the source term $\bf{f}$ is a random 
large-scale driving force\footnote{${\bf f}= \rho d{\bf v}/dt$}.   
We drive turbulence solenoidally, at wave scale $k$ equal to about 2.5 
(2.5 times smaller than the size of the box). This scale defines 
the injection scale in our models in Fourier space to minimize 
the influence of the forcing on the generation of density structures. 
Density fluctuations are generated later on by the interaction of MHD waves. 
%We use units in which $V_A=B_\mathrm{ext}/\left( 4 \pi \rho \right)^{1
%/2}=1$. The average RMS velocity $\delta V$ in a statistically stationary state is around $1$.
%The RMS velocity $\delta V$ is maintained to be approximately unity, 
%so that ${\bf v}$ can be viewed as the velocity measured in 
%units of the RMS velocity of the system and ${\bf B}/\left( 4 \pi \rho \right)^{1/2}$ 
%as the Alfv\'{e}n velocity in the same units. 
The time $t$ is in units of the large eddy turnover time 
($\sim L/\delta V$) and the length in units of $L$, the scale 
of energy injection.
The scale of energy dissipation is  defined by the numerical 
diffusivity of the scheme\footnote{The ENO-type schemes are considered to be relatively 
low diffusion ones \cite[see][e.g.]{liu98,levy99}. The numerical diffusion depends 
not only on the adopted numerical scheme but also on the ``smoothness'' of the solution, 
so it changes locally in the system. In addition, it is also a time-varying quantity.}.
The magnetic field consists of the uniform background field and a 
fluctuating field: ${\bf B}= {\bf B}_\mathrm{ext} + {\bf b}$. Initially ${\bf b}=0$.

We divided our models into two groups corresponding to 
sub-Alfv\'enic ($B_\mathrm{ext}=1.0$) and 
super-Alfv\'enic ($B_\mathrm{ext}=0.1$) turbulence. 
For each group we computed several models with different values of 
gas pressure (see Table \ref{tab:models}). 
We ran compressible MHD turbulent models, with 512$^3$ resolution, 
for $t \sim 5$ crossing times, to guarantee full development of energy cascade. 
Since the saturation level is similar for all models and we solve the isothermal MHD 
equations, the sonic Mach number is fully determined by the value of the isothermal 
sound speed, which is our control parameter.  
The models are listed and described in Table~\ref{tab:models}.

\begin{table}
\begin{center}
\caption{Description of the simulations - MHD, 512$^3$ 
\label{tab:models}}
\begin{tabular}{cccccc}
\hline\hline
Model & $p_{gas}$ & $B_{\rm ext}$ & ${\cal M}_s$ & ${\cal M}_A$ &Description \\
\tableline
1 &2.00 &1.00 &0.1 &0.7 & subsonic \& sub-Alfv\'enic     \\
2 &1.00 &1.00 &0.7 &0.7 & subsonic \& sub-Alfv\'enic     \\
3 &0.10 &1.00 &2.0 &0.7 & supersonic \& sub-Alfv\'enic   \\
4 &0.025 &1.00 &4.4 &0.7 & supersonic \& sub-Alfv\'enic   \\
5 &0.01 &1.00 &7.0 &0.7 & supersonic \& sub-Alfv\'enic   \\
6 &0.0077 &1.00 &8.4 &0.7 & supersonic \& sub-Alfv\'enic   \\
7 &0.0049 &1.00 &10 &0.7 & supersonic \& sub-Alfv\'enic   \\
8 &1.00 &0.10 &0.7 &2.0 & subsonic \& super-Alfv\'enic   \\
9 &0.10 &0.10 &2.0 &2.0 & supersonic \& super-Alfv\'enic \\
10 &0.025 &0.10 &4.4 &2.0 & supersonic \& super-Alfv\'enic \\
11 &0.01 &0.10 &7.0 &2.0 & supersonic \& super-Alfv\'enic \\
12 &0.0077 &0.10 &8.4 &2.0 & supersonic \& super-Alfv\'enic   \\
13 &0.0049 &0.10 &10 &2.0 & supersonic \& super-Alfv\'enic   \\
\hline\hline
\end{tabular}
\end{center}
\end{table}

\begin{figure*}[tbh]
\centering
\includegraphics[scale=.3]{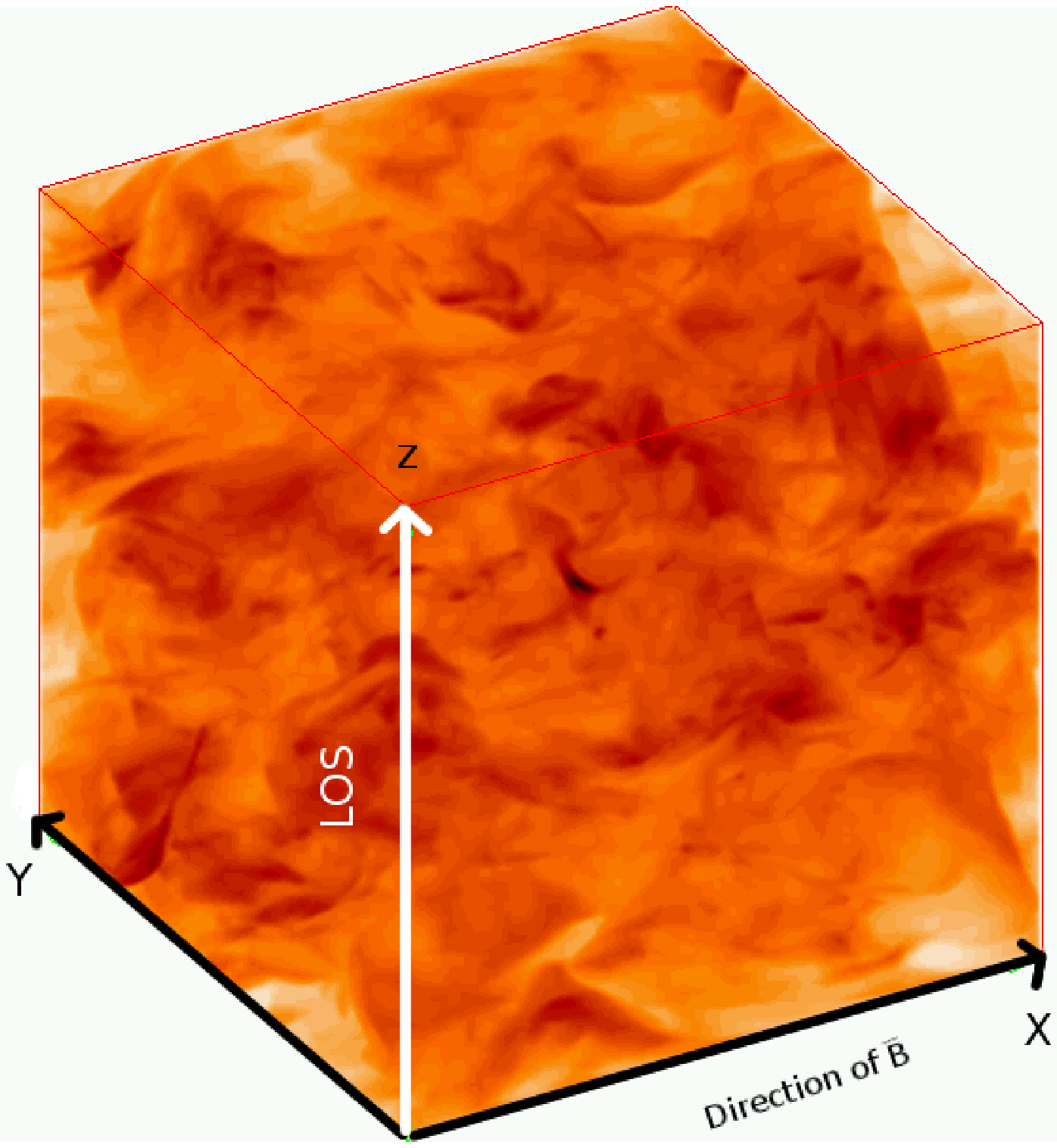}
\includegraphics[scale=.25]{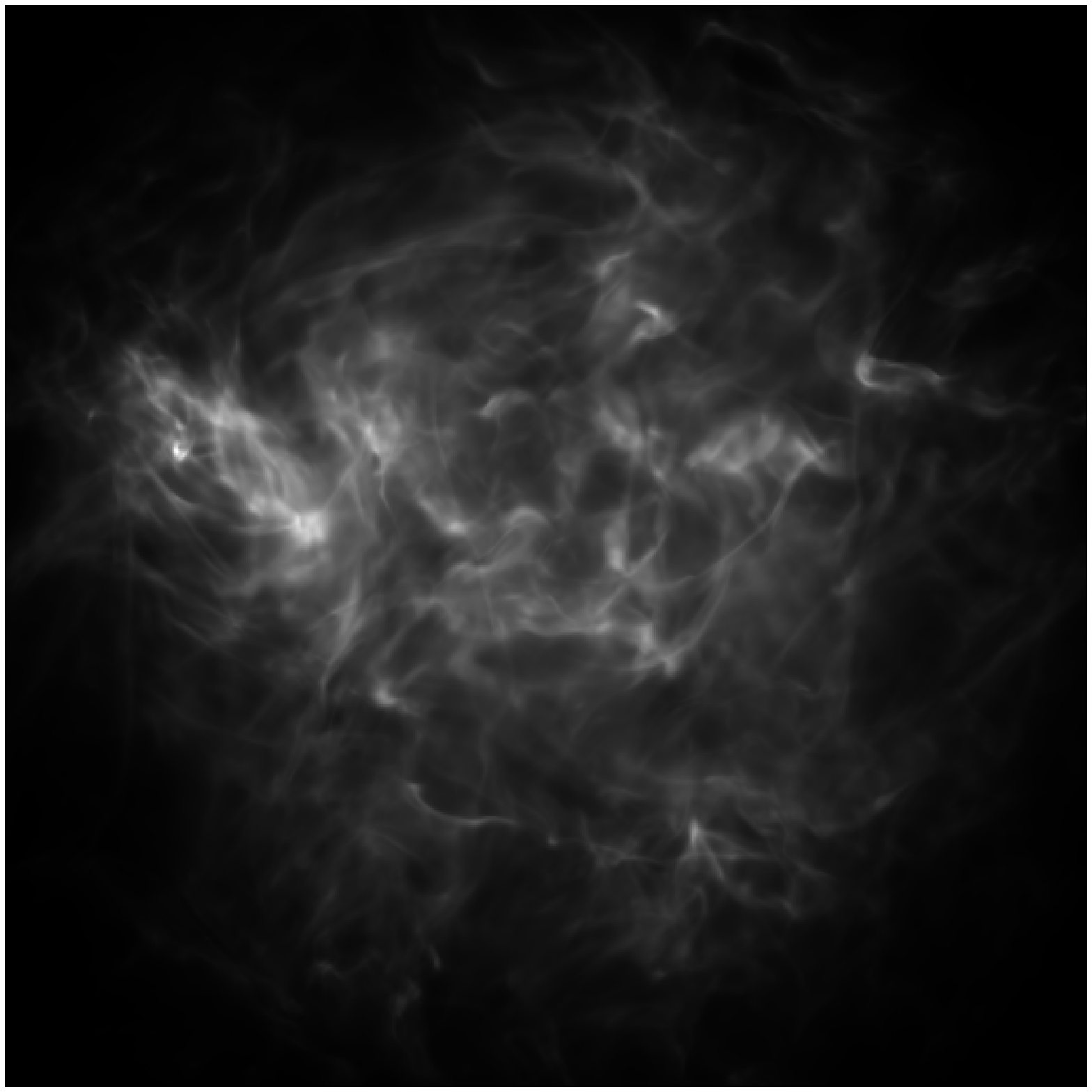}
\caption{Left:  
An example of a simulated subsonic, sub-Alf\'venic (${\cal M}_s=0.7$, ${\cal M}_A=0.7$)  
$512^{3}$ density data cube demonstrating one possible direction of the LOS used to 
calculate the column density distribution. 
The mean magnetic field is oriented along the x direction.
Here the LOS is along the z axis, although throughout the paper we integrate 
density fields along all three axes.
Right: A supersonic column density image of a simulation with cloud boundaries, 
with a boundary radius similar (in pixel size) to what is seen for the SMC.}
\label{fig:fakecloud}
\end{figure*}

For each model, the results of MHD simulations are: the isothermal 3D density field, 
three components of velocity ($V_x$, $V_y$, $V_z$),
and three components of magnetic field. 
As an example,  Figure~\ref{fig:fakecloud} (left) shows a density  field for a sub-Alfv\'enic, subsonic
simulation. To calculate the column density distribution we integrate the 3D density
fields along a given LOS.
We introduce the following nomenclature  throughout the paper:
``x column density" or ``column density in the x direction"  refers to the density cube being 
integrated along the x direction (parallel to the mean magnetic field).  
This description is similar for the y and z directions (perpendicular to $B_{ext}$). 
In the case of the Figure~\ref{fig:fakecloud} (left), the LOS is along the z axis,
and the magnetic field is oriented along the x axis.

\subsubsection{Scaling of the column density}

In order to compare simulations with the SMC data, we apply a simple scaling
to the simulated column density distributions:
\begin{equation}
N_{norm}=\frac{N-\langle N \rangle}{\sigma(N)},
\end{equation}
where $\sigma(N)$ denotes the standard deviation  and $\langle N \rangle$ denotes the 
arithmetic  mean of the column density image. This method, often referred to as the
 \textit{standard score}, standardizes all data sets used in this study.
After the scaling, the new data set has values which represent
the difference between the original data values and the sample mean, in units of 
the standard deviation. 
The same scaling was applied on the SMC HI column density image.

%The scaling of simulations to match a particular physical situation has been done in 
%many previous works \cite[see][for example]{Padoan03,Hill08}. In this study we deal 
%only with statistics that are effected by density, and hence we are only concerned 
%with scaling densities. 

\subsubsection{Simulating cloud boundaries}

Another consideration to take into account is the fact that simulated cubes are periodic and 
do not have boundaries. In other words, simulations do not 
have a decrease in the column density values that is seen in the SMC data 
as one goes radially out from the center of the image.  
KLB showed that boundaries affect higher statistical moments of density, 
but have little effect on the spatial power 
spectrum and the bispectrum, generally only impacting the large scales.

To introduce cloud boundaries to our simulations, we
multiply the simulated 3D density fields by a 3D spherical function, which has a value of one
within a sphere of radius $R$  and is decaying outside of this radius with a Gaussian profile. 
We choose a $R$ value of 128 pixels,
which is  4 times smaller then the box size.
An example of the simulation with a boundary is shown in the right-hand panel 
in Figure~\ref{fig:fakecloud}.  
Please note that we impose cloud boundaries on the simulated cubes of homogeneous
turbulence after turbulence has been fully developed.
This results in the simulated column density profiles being similar
to the SMC profile, however, is different from simulations of turbulence
being developed within cloud boundaries.

%We stress that our simulations are isothermal, have no preferred spatial scale, and 
%the box does not correspond to any specific physical size.  
%The lower limit on the range of scales is set only by the resolution of the simulation.

\section{Higher Order Moments: Simulated data}

\label{sec:moments}

The most straightforward statistical properties of a distribution
are the mean value and the variance. The mean arithmetic value
and the variance of the distribution of, for example, column density $\xi$ are given by:
\begin{equation}
\bar{\xi}=\frac{1}{N}\sum_{i}^N \xi_i
\end{equation}
and 
\begin{equation}
\sigma_{\xi}^2=\frac{1}{N-1}\sum_{i}^N (\xi_i-\bar{\xi})^2
\end{equation}
where $N$ is the number of samples or points of the mesh in the case of simulation data.
The mean value is a less important property in our studies
so we do not consider it here;
however, it is required to calculate higher moments. Variance measures the width of the
distribution and is always posit%\begin{figure*}[tbh]
%\centering
%\includegraphics[scale=.42]{f11a.eps}
%\includegraphics[scale=.42]{f11b.eps}
%\caption{OLD:Contours of sonic Mach number in the SMC calculated from statistical moments. 
%HI is plotted on the left and H $\alpha$ on the right. Minimum values of ${\cal M}_s$ lie 
%close to zero while the max is approximately 10. The highest values of ${\cal M}_s$ 
%are along the bar and east wing.}
%\label{fig:ms_var}
%\end{figure*}ive by definition.  Skewness and kurtosis are defined 
by the third and fourth-order statistical moment. Skewness is defined as:

\begin{equation}
\gamma_{\xi} = \frac{1}{N} \sum_{i=1}^N{ \left( \frac{\xi_{i} - \overline{\xi}}{\sigma_{\xi}} \right)^3 }.
\label{eq:skew}
\end{equation}
If a distribution $\xi$ is Gaussian, its skewness is zero. Negative skewness indicates 
the data are skewed in the left direction (the tail of the PDF is extended to the left, or
towards low density values), while positive values imply that the distribution is skewed 
in the right direction (the tail of the PDF is extended to the right, or towards high density values).

Kurtosis is a measure of whether a quantity has a distribution that is peaked or 
flattened compared to a Gaussian distribution.  Kurtosis is defined in a 
similar manner to skewness, only is derived from the forth order statistical moment.  
If a data set has positive kurtosis then the PDF of data values  
will have a distinct sharp peak near the 
mean and elongated tails.  If a data set has a negative kurtosis then its PDF will be 
flat at the mean. Kurtosis is defined as:

\begin{equation}
\beta_{\xi}=\frac{1}{N}\sum_{i=1}^N \left(\frac{\xi_{i}-\overline{\xi}}{\sigma_{\xi}}\right)^{4}-3.
\label{eq:kurt}
\end{equation}

\begin{figure*}[tbh]
\centering
\includegraphics[scale=.5]{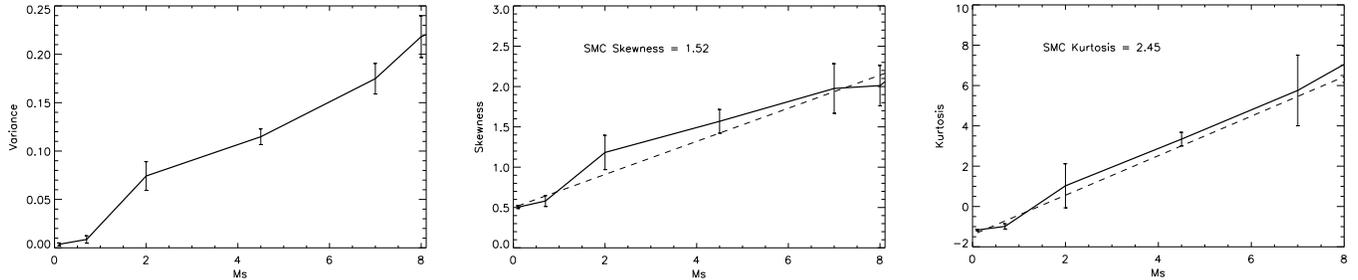}
\caption{Variation of the variance (left), skewness (middle), and kurtosis (right) with the sonic 
Mach number derived from the final snapshot of each simulation with cloud boundaries applied. 
For all simulations ${\cal M}_A$=0.7.
Plotted data points are averages from data sets integrated along different LOS, 
with error bars representing the standard deviation between column density images with 
different LOS.  
As skewness and kurtosis increase almost linearly with  ${\cal M}_s$, a linear function 
was fit over the range ${\cal M}_s\sim 1-8$ and is shown with a dashed-line.
The measured skewness and kurtosis of the SMC HI column density
image are: skewness$=1.52 \pm0.01$ and kurtosis$= 2.45\pm0.02$.  
Both of these values suggest that the HI in the SMC
is supersonic, with a sonic Mach number $\geq 3 $ within the error bars. 
This is discussed in Section 5.
}
\label{fig:kowal}
\label{fig:skewplot}
\end{figure*}

Variance, skewness and kurtosis of simulated column densities depend on turbulent properties, 
specifically the sonic Mach number.  
This is shown in Figure~\ref{fig:kowal} for
the simulated column density distributions with cloud boundaries.
Error bars are computed by estimating the standard deviation between models with the
same sonic Mach number but differing LOS orientations.
All three moments depend almost linearly on the sonic Mach number,
for ${\cal M}_{s} \ga 1-2$. KLB noticed a similar trend for supersonic simulations 
but focused only on  ${\cal M}_{s}=0.2-2$. Our simulations extend this result all
the way to ${\cal M}_{s}=8$. 
For ${\cal M}_{s} \la0.5$ KLB found that both 
skewness and kurtosis have a rather flat dependence on the sonic 
Mach number, while the variance continues to decreases gradually for subsonic models.
The increase of higher order moments with ${\cal M}_{s}$ can be explained 
in that as the Mach number increases, 
the abundance of small-scale density structure increases, resulting
in a broader density PDF. 
BFKL showed that the increase of higher-order moments with ${\cal M}_{s}$ is not a resolution
effect, since they replicated the result of 
KLB, (who used cubes of 128$^3$ resolution) with cubes of 512$^3$ resolution.

Which one of three high-order moments represents the best statistics to 
describe turbulence in observational data?
While variance has a linear dependence over a broad range of ${\cal M}_{s}$ values,
it depends on the exact  scaling of the data set, making
a direct comparison between simulations and observations difficult.
On the other hand,
skewness and kurtosis are, by definition, normalized by the standard deviation
and do not have this problem. 
In addition, variance changes with the length along the LOS, since perturbations can add up in a
random walk fashion.  
All of this, as well as the result by KLB that kurtosis is the least affected by cloud boundaries,
suggest that for our immediate comparison of observations 
with the MHD simulations, the higher order moments, skewness and kurtosis, 
are more appropriate statistics.

\subsection{Spatial Distribution of Skewness and Kurtosis}
\label{sec:mom}

While it is useful to know the Mach number of the ISM in a global sense, even more 
interesting is to see how it varies spatially and whether it correlates with 
local star forming regions where high turbulence is expected.  
To characterize small scale departures from Gaussian distributions 
within the column density distributions, we create moment maps of the 
simulated column densities using a circular moving kernel. We do this by moving a circle of a given 
radius $r$ across the image and calculating  the skewness and kurtosis at all points. 
\begin{figure*}[tbh]
\centering
\includegraphics[scale=.5]{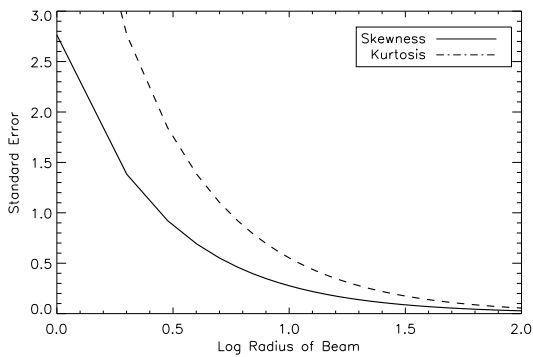}
\caption{The standard error in skewness (SES) and kurtosis (SEK) vs. radius of the circular kernel
(related to the number of points or pixels within the kernel used to calculate the 
skewness/kurtosis).  SES$=\sqrt{6/N}=\sqrt{\frac{6}{\pi r^2}}$, 
while SEK$=\sqrt{24/N}=\sqrt{\frac{24}{\pi r^2}}$.  
We choose $r=35$ pixels to retain good resolution and low SEK/SES.
These error measurement describes how large the statistics must be before they are 
deemed significant deviations from Gaussianity. 
If statistics are in this range 
then they are deemed generally Gaussian.  If statistics are
outside of this range then departures from Gaussianity have occurred.}
\label{fig:skewerror}
\end{figure*}

First we must decide on a kernel size that will enclose enough pixels to provide reliable statistics. 
According to Tabachnick \& Fidell (1996), the standard error in skewness (SES) can be 
estimated by $\sqrt{6/N}$ and the standard error in kurtosis (SEK) can be estimated
by $\sqrt{24/N}$, where $N$ is the number of points used to calculate the skewness/kurtosis. 
Generally, if a value of skewness/kurtosis falls within $\pm 2 \times $SES/SEK, then the distribution
is considered to be normal, while values of skewness/kurtosis larger than twice the absolute value of 
SES/SEK imply significant non-Gaussianity. Figure ~\ref{fig:skewerror} shows 
$2\times$ SES and SEK as a function of the kernel radius $r$.  
Clearly, SES/SEK is proportional to $1/r$; the smaller the radius, the higher 
the absolute value of SES/SEK.  Thus, we select $r=35$ pixels, which corresponds 
to the point where $2\times$ SES/SEK starts to deviate from zero and 
ensures that Gaussianity is 
represented by  $2\times$ SES/SEK$\sim0$.

Using a circular moving kernel we generate 3rd and 4th moment maps of
simulated column density distributions. 
To ensure there are no resolution artifacts,
we briefly explore correlations between skewness and kurtosis
of the derived moment maps on a pixel-to-pixel basis.
Based on the work by KLB and BFKL, and also our Figure~\ref{fig:kowal}, we expect for 
supersonic models a correlation of both skewness and kurtosis with 
the sonic Mach number, and that skewness and kurtosis agree in sign and relative value. 
However, as subsonic models show unexpected 
behavior at low resolutions (from the KLB study), the dependence of 
skewness and kurtosis on the sonic Mach number is not 
strong. 

\begin{figure*}[tbh]
\centering
\includegraphics[scale=.6]{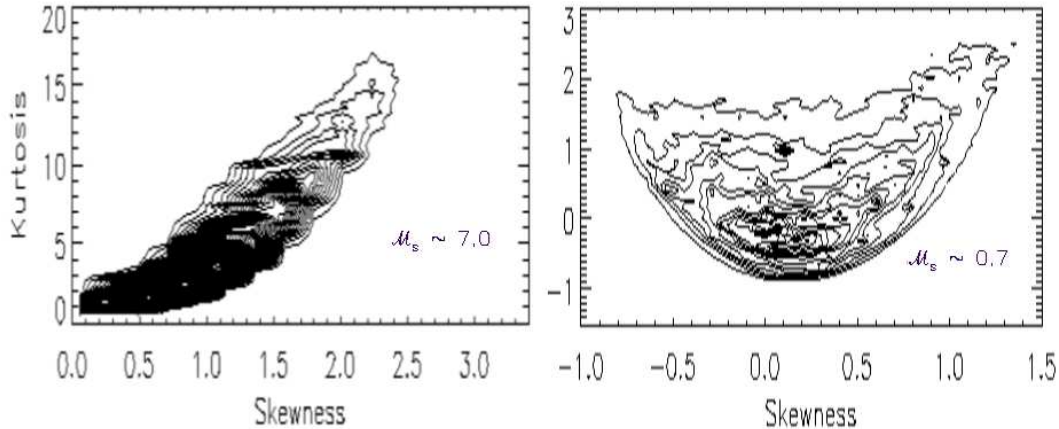}
\caption{Kurtosis vs. Skewness for isothermal simulations with a 
kernel of $r=35$ pixels.  The left is a supersonic model,
while the right is a subsonic model. The supersonic model shows good correlation 
between skewness and kurtosis while the subsonic model does not. }
\label{evidence}
\end{figure*}

Figure~\ref{evidence} shows pixel-to-pixel comparison between 
skewness and kurtosis for two example models: a supersonic with  ${\cal M}_{s}=7.0$ 
(left panel) and a subsonic with ${\cal M}_{s}=0.7$ (right panels).
It is evident that the supersonic model shows a good correlation between 
kurtosis and skewness, while the subsonic model does not.
This is exactly as expected. This result
proves that our approach of deriving moment maps is not resolution limited
although we use a smaller number statistics within each circular kernel
($\pi r^2=1380$ pixels).
This also highlights another interesting property: regions of moment
maps with a good correlation between skewness and kurtosis could be
interpreted by supersonic MHD turbulence, while
a poor correlation may be caused by subsonic turbulence.
We present in the appendix another technique that 
could be used to further identify subsonic regions.

\section{Higher order moments: the SMC HI column density image}
\label{sec:moments-smc}
\label{sec:mmaps}

We first calculate the higher order moments of the whole SMC HI column density image. 
We find that the skewness is 1.52$\pm0.01$, and the kurtosis is 
$2.45\pm0.02$. The estimated uncertainties are determined by 
calculating twice the SES or SEK. 
Figure~\ref{fig:skewplot} suggests that on average the HI in the SMC is supersonic
with ${\cal M}_{s} \ga3$.

To characterize small scale departures from the Gaussian distribution within the SMC 
we create 3rd and 4th moment maps of the SMC using a circular kernel, 
as discussed for simulations 
in Section~\ref{sec:mom}.  We repeat a similar process for the SMC as we did with 
the simulations.  We use a radius of $r=35$ and calculate moments 
in a circular kernel moving across the SMC.
This is equivalent to convolving the original HI image with
a Gaussian function with a FWHM of 30$'$. The resultant
moment images therefore have a lower resolution than the original HI image.  
 After this we 
create a mask which cuts off pixels below a column density of 
$1.26\times10^{21}$ cm$^{-2}$ in the HI column density image, 
to exclude increase in noise along the edges of the image. 
This is done to eliminate noisy pixels in moment maps, since pixels along the 
edge have significantly smaller column density values.

\begin{figure*}[tbh]
\centering
\includegraphics[scale=.45]{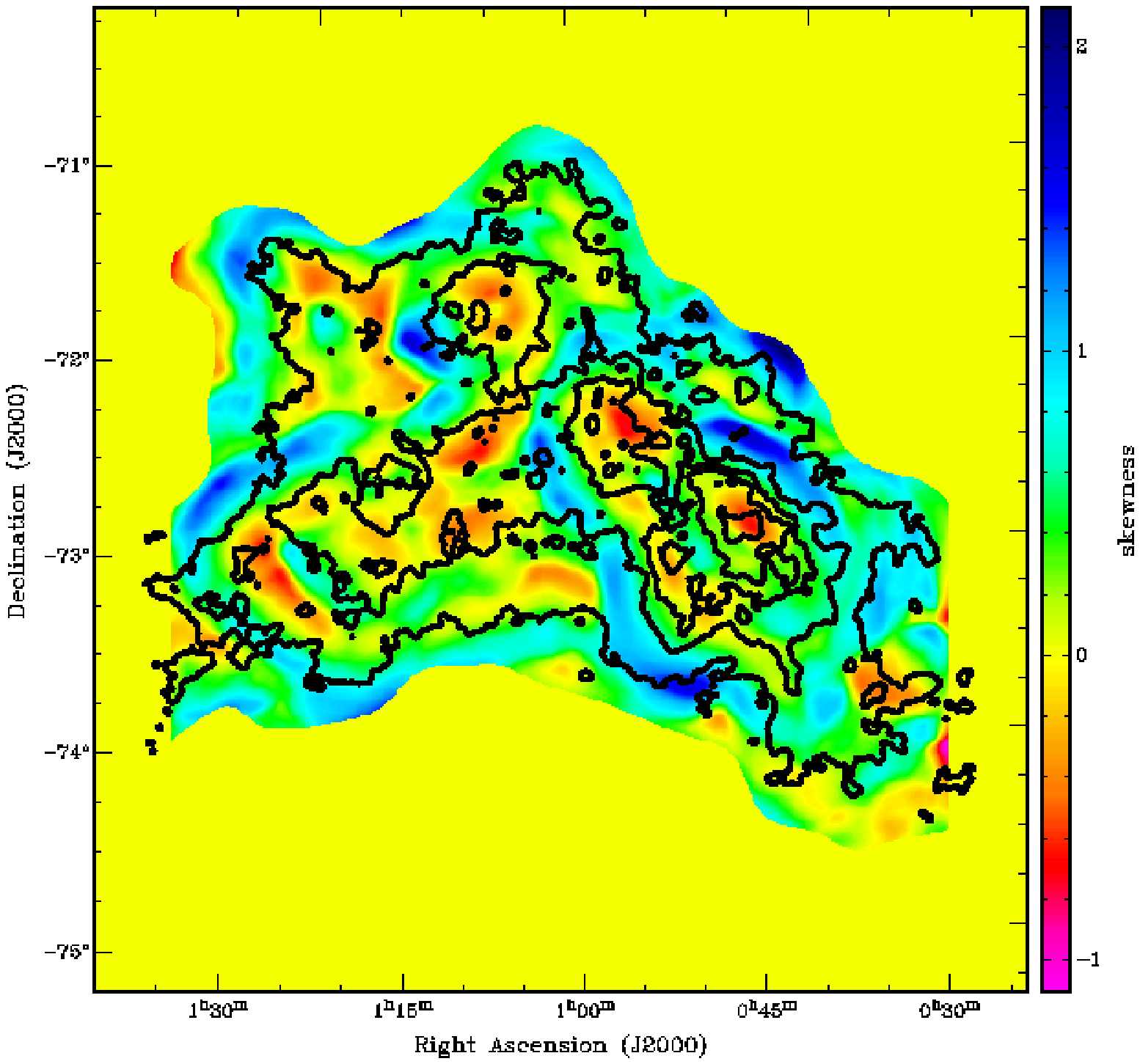}
\includegraphics[scale=.45]{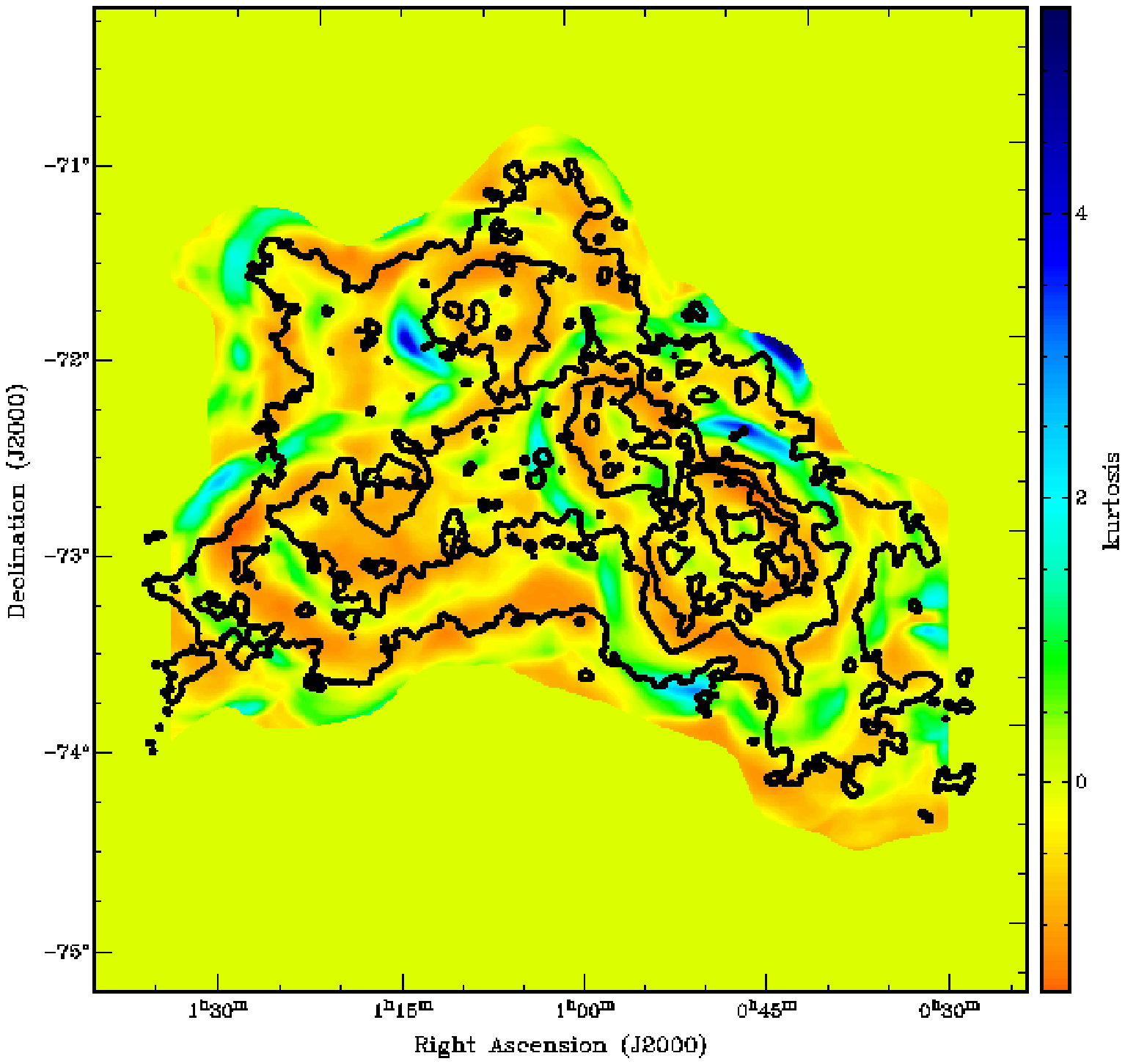}
\caption{ (Left)
Skewness of the HI column density derived using a circular kernel with $r=35$ pixels.  
HI contours are overlaid.
(Right) Kurtosis of the HI column density with the HI contours overlaid. 
The contours show the HI column density and range from 20 to 90\%, with a step of 15\%.}
\label{fig:skewmap}
\end{figure*}

Figures~\ref{fig:skewmap} 
shows the distribution of skewness and kurtosis across
the SMC (the HI column density image was standardized using the standard score
method), with overlaid HI contours smoothed to 30$'$ resolution.
These maps  retain the overall shape of the SMC, which can be seen in the HI 
contours. Familiar features of the SMC, such as the east wing and bar, regions of 
high star formation, can be picked out in these maps.

Generally, skewness ranges from $-0.5$ to 1, with  several discrete 
regions reaching higher
positive or negative values. For kurtosis, most pixels range from $-1$ to 1, with again
several exceptions. Based on Figure 3, this suggest that large areas in the SMC have
${\cal M}_{s}= 0-2$ and are subsonic or transonic.
Along most of the bar the skewness and kurtosis correlate well.
This suggests that the MHD turbulence could be the cause of local
departures from Gaussianity in the HI column density distribution.
We warn once again that both our study and earlier work
find strong dependence of skewness and kurtosis on the sonic Mach number only for
supersonic models, while the dependence
for subsonic models is rather weak. 
Interpreting subsonic turbulence is therefore difficult, and poor correlation
between skewness and kurtosis could be due to either subsonic turbulence
or some other physical mechanisms.
Several interesting regions stand out in Figure~\ref{fig:skewmap}.

\begin{itemize}

\item Along the HI bar and the Eastern Wing region
both skewness and kurtosis are negative, reaching
values even lower than what is seen in Figure 3 for global averages.
Based on expectations from numerical simulations this
could be  explained by the subsonic isothermal MHD turbulence,
as shown in Figure 5. 
Most of the HI bar and the Wing therefore could be
explained as being subsonic, with several
discrete concentrations with almost no turbulence, probably tracing
quiescent regions
and potential sites of future star formation.
However, as subsonic regions are hard to constrain from simulations, 
additional processes may be also playing an important role.

\item From HI peaks radially outward, both skewness and kurtosis 
gradually increase.
The highest values of both skewness ($\sim2$) and kurtosis ($\sim3-4$) 
are found along the HI bar and correspond to
areas of compressed HI contours. These regions could be interpreted as  
having the highest level of turbulence (with ${\cal M}_{s}= \sim4$) in the SMC. 
This suggests that the most turbulent regions may be associated with 
the shearing flows and/or shocks between the bar and the surrounding HI.

\item Sudden change in the behavior of both skewness and kurtosis can be 
noticed around RA 01$^{\rm h}$ 10$^{\rm m}$, Dec $-72^{\circ}$ 10$'$.
For example, skewness flips from high positive to high negative
values over an angular scale of $\sim60'$. Interestingly, 
this flip happens in the direction
of an HI extension towards the LMC and again may be pointing out
streaming or tidal motions caused by the interactions between
the SMC and the LMC. 

\end{itemize}

\begin{figure*}[tbh]
\centering
\includegraphics[scale=.5]{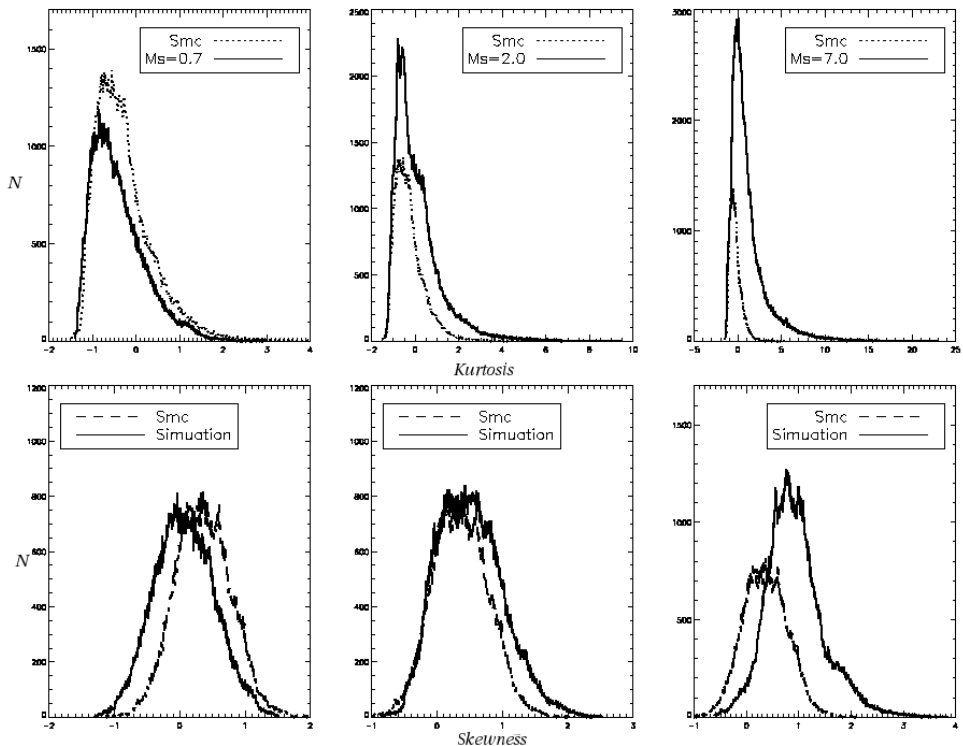}
\caption{Histograms of the skewness and kurtosis maps from the simulations and the SMC 
HI column density image 
(see Figure~\ref{fig:skewmap}), derived using the moving 
circle kernel with $r=35$.  The top row shows kurtosis, and the bottom row shows skewness. 
The SMC distributions are overplotted with dashed lines and agree  
best within the ${\cal M}_{s}= 2.0$ simulations.}
\label{fig:kurtsmcsim}
\end{figure*}

% Compare skew/kurtosis range of values with simulations:

In order to compare the range of values found in the SMC and the simulations 
we plot several histogram of the skewness and kurtosis values 
in Figure~\ref{fig:kurtsmcsim}. These figures 
show that the majority of SMC pixels generally match well with the mildly 
supersonic models with ${\cal M}_{s}\sim1-2$.  
However, the SMC distribution appears to be more narrow in terms of both
skewness and kurtosis values than the ${\cal M}_{s}\sim2$ simulation.
For subsonic and very supersonic models, the values of skewness and kurtosis of the SMC 
are not in the range  of  the simulations.

\begin{figure*}[tbh]
\centering
\includegraphics[scale=.45]{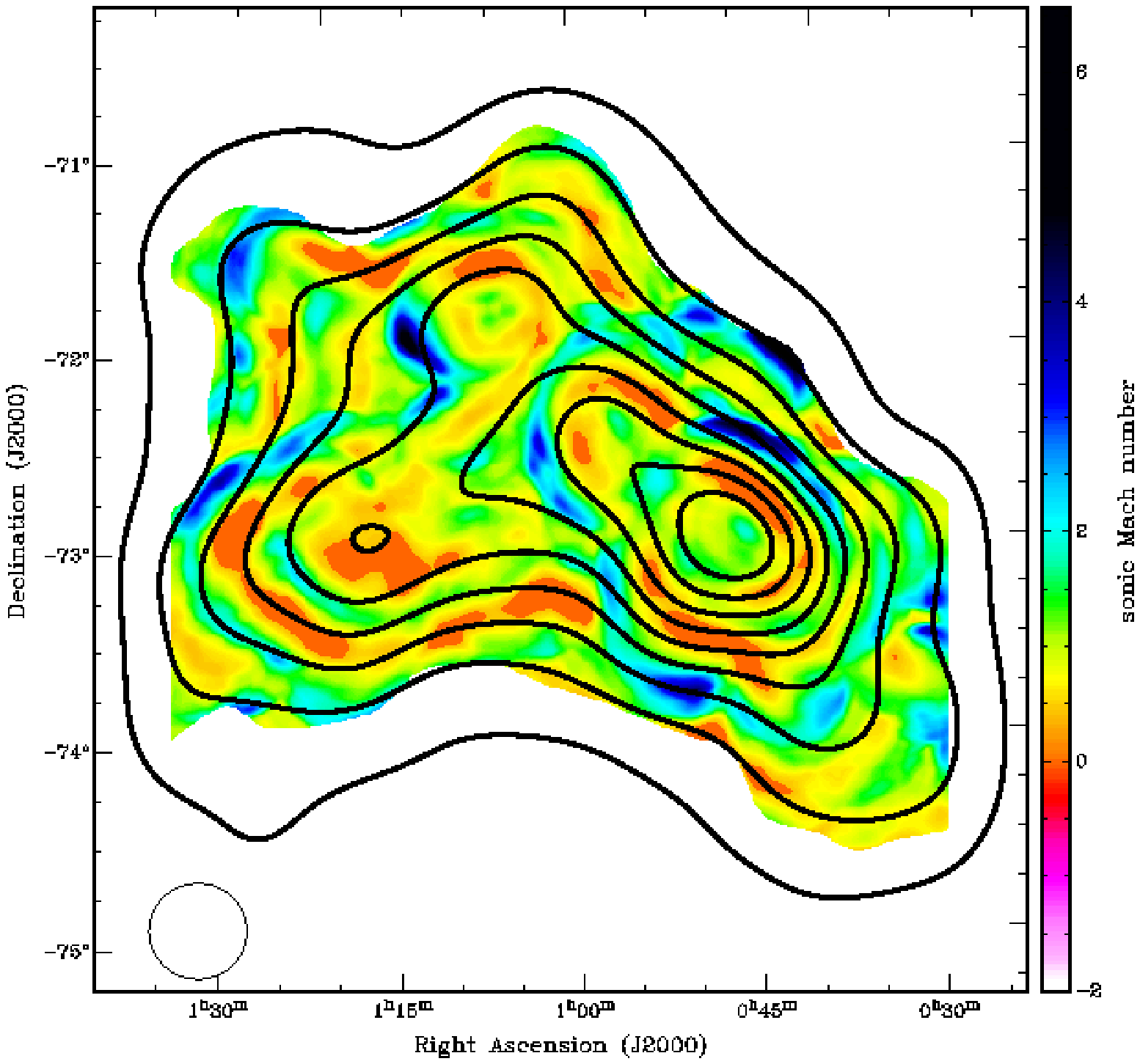}
\includegraphics[scale=.45]{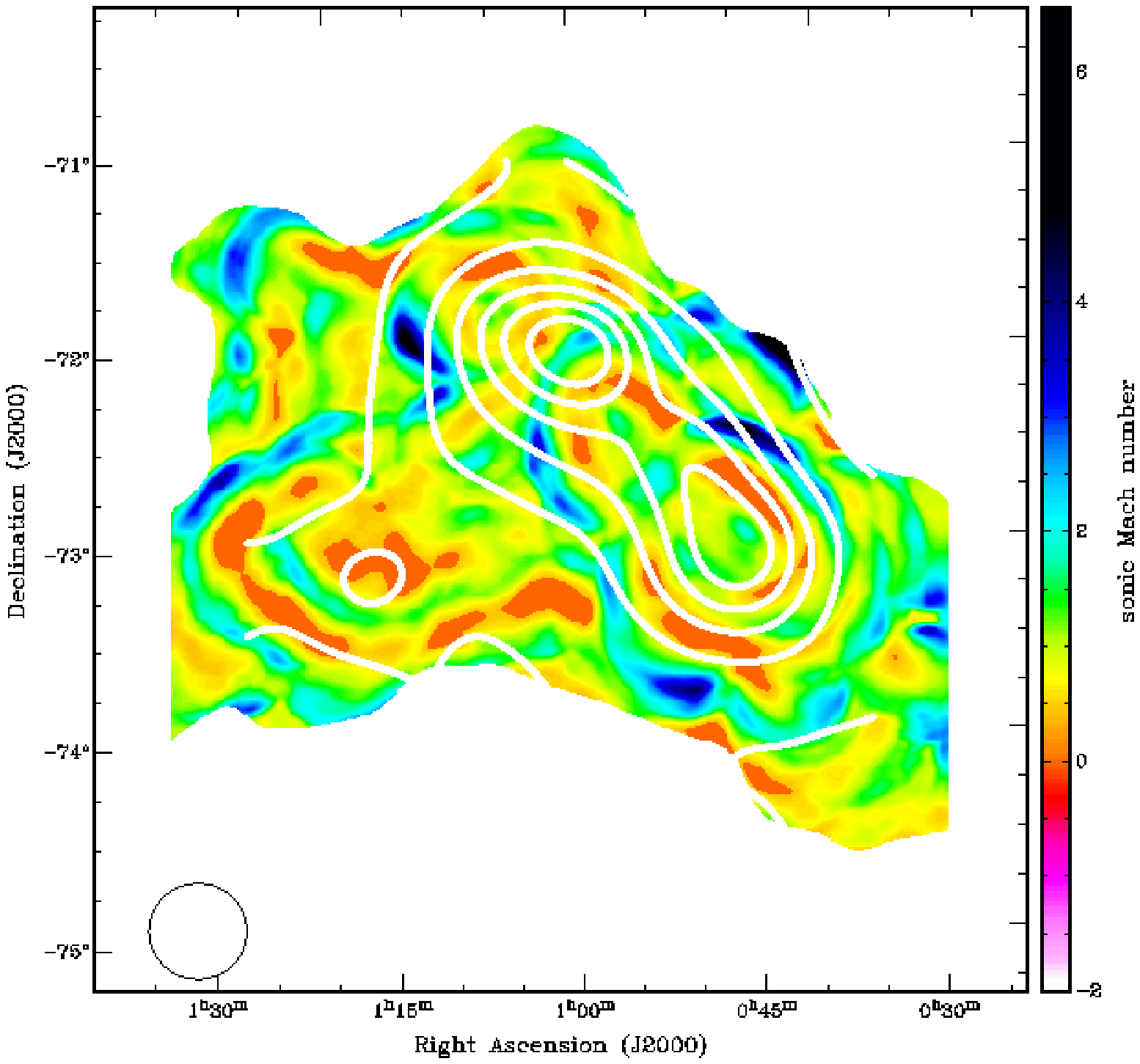}
\caption{ (Left) The sonic Mach number image derived 
from the HI column density image of the SMC overlaid with the HI column density
contours. The circle in the bottom-left shows the angular resolution of the image. 
(Right) Same as on the left but overlaid with the H$\alpha$ contours smoothed to
the same resolution. The H$\alpha$ image is from Kennicutt et al. (1995).
}
\label{fig:ms_var}
\end{figure*}

\subsection{The ${\cal M}_{s}$ map of the SMC}

As shown above, with our limited angular resolution
of 30$'$ we see strong hints that different regions in the SMC have different 
turbulent properties, suggesting
spatial variations of ${\cal M}_{s}$. We would now like to combine
all this information into an image of ${\cal M}_{s}$ for the SMC.
As already discussed, 
based on the KLB study, kurtosis is the least prone to effects caused by
cloud boundaries and as Figure 3 suggests
has an almost linear shape for ${\cal M}_{s}\sim1-8$.
Therefore, we fit a linear function for
all values of kurtosis $>\sim-1$. This function can be inverted to
estimate ${\cal M}_{s}$:
${\cal M}_{s} \sim ({\rm kurtosis} + 1.44)/1.05$.
Pixels in the ${\cal M}_{s}$ image with corresponding kurtosis 
$<-1$ are set to zero and most likely mark subsonic
regions, or could be caused by additional processes.
The resultant map is shown in Figure~\ref{fig:ms_var} (left).

% Interpretation of the Ms map: 
As expected already,
most of the area of the SMC bar and the Eastern Wing 
could be interpreted as having subsonic or transonic 
Mach numbers, ${\cal M}_{s}\sim0-2$.
Several concentrated regions across the SMC indicate
very quiescent environments, most of them unfortunately have a size close
to our angular resolution.
Regions with the highest sonic Mach number, ${\cal M}_{s}\sim4$, are 
found around the bar and correspond to compressed HI contours. 

We can quantify the fraction of HI with different ${\cal M}_{s}$.
The most likely quiescent regions (set to zero in our map) 
comprise 8\% of the mapped area.
Regions with $0<{\cal M}_{s}\le1$ comprise about 40\%,
while about the same fraction is contained in transonic regions 
with ${\cal M}_{s}=1-2$.
Regions with higher turbulence, ${\cal M}_{s}>2$ constitute about 10\% of the area.

% Comparison with Halpha:
Figure~\ref{fig:ms_var} (right) shows contours of the H$\alpha$ emission (obtained by
Kennicutt et al. 1995) smoothed to 30$'$ and overlaid on the ${\cal M}_{s}$ map.
Sites of the most recent star formation in the bar and the Eastern Wing 
have ${\cal M}_{s}\sim1$, suggesting that
the most turbulent regions are
not associated with star formation. The most turbulent regions appear to trace
shearing and tidal motions.
We note though that our resolution of 30$'$ is too low to trace individual
 star-forming regions.

% Comparison with vel. dispersion map:
As an estimate of the velocity dispersion along the LOS is often
used as a possible measure of HI turbulence,
we investigated how the HI velocity dispersion compares to 
our ${\cal M}_{s}$ map. 
Stanimirovic et al. (2004) found that  regions
with the highest dispersion ($\sim30$ \kms) appear to be associated with the positions
of the three largest super-giant shells.
We find no obvious correlation between the HI velocity dispersion and our ${\cal M}_{s}$ map.

This lack of correlation could be due to projection effects, and/or 
large LOS depth of the SMC. In addition, velocity dispersion (similar to velocity centroids) is subject to 
the interplay of several statistical effects (Lazarian \& Esquivel 2003, 
Esquivel \& Lazarian 2005, Esquivel et al. 2007). 
The most obvious is the contribution of both velocities and densities to the 
resulting measure. More subtle, but still important, are effects of velocity-density 
correlations and non-Gaussianity. These effects make centroids unreliable when dealing 
with supersonic turbulence (Esquivel et al. (2007). Effects of phase transitions resulting in the existence of
cold and warm HI and the pliable equation of state corresponding to interstellar HI are
likely to act in a similar way. Thus the velocity dispersions obtained from 
spectral lines  are substantially affected by densities, 
which are the quantities that we deal in this study.
They should be distinguished from the true velocity dispersions, which, unfortunately, are not 
available through spectral line observations.

\section{Comparison with the observationally inferred ${\cal M}_{s}$ of the CNM}
\label{ratio}

Constraining the sonic Mach number of the warm neutral medium (WNM) 
observationally is highly difficult
due to the lack of direct measurements of the HI temperature.
This is the main reason why in the previous section 
we used the HI column density image of the SMC and a database of
numerical simulations to explore spatial variations of ${\cal M}_{s}$.
To investigate reliability of our results we compare our derived
${\cal M}_{s}$ distribution  with results from an observational method 
which constrains temperature of the cold HI, and therefore ${\cal M}_{s}$ of the CNM.
This method is based on 
the comparison of the spin temperature ($T_s$) with the upper limit
on the kinetic temperature ($T_{k,max}$) for CNM clouds seen in HI absorption.

For observational data we use HI absorption and emission 
observations of 29 radio continuum sources behind the SMC by Dickey et al. (2000).
For each of the sources \cite{Dick00}
provided the FWHM linewidth of the absorption spectra, and
estimated the spin temperature of the CNM seen in absorption.
By assuming Doppler broadening of HI velocity profiles, we can estimate the upper
limit on the kinetic temperature as $T_{k,max} =(FWHM/0.215)^2 $.
%The difference between the measured $FWHM$ and the expected thermal linewidth
%at a given spin temperature, provides an estimate of the turbulent velocity. 
As shown in \cite{Heiles03}, the ratio of  
$T_{k,max}$ and the spin temperature, $T_s$, is related to
the 1D mean square turbulent velocity: 
\begin{equation}
 V_{t,1D}^{2}=\frac{kT_s}{m_{H}}\left( \frac{T_{k,max}}{T_s}-1\right), 
\end{equation}
where $k$ is the Boltzmann constant and $m_H$ is the mass of the hydrogen atom.  
Multiplying this by 3 gives the mean square 3D turbulence 
velocity $V_{t,3D}^{2}$ of the CNM.  To estimate the 
sonic Mach number we need to 
divide $V_{t,3D}$ by the sound speed $C_s=\sqrt{kT_s/ \mu m_H}$.
We adopt a mean atomic 
weight of $\mu=1.22$ for the ISM of SMC abundance (see Mao et al. 2008).
Using this, we can write: 

\begin{equation}
{{\cal M}_s}^2= \frac{{V_{t,3D}}^2}{{C_s}^2}=3.7\left( \frac{T_{k,max}}{T_s}-1\right) .
\end{equation}

For each of 29 radio continuum sources from \cite{Dick00}
we derive $T_{k,max}$ and use the estimated $T_s$ to
calculate the sonic Mach number of the CNM clouds.
\cite{Dick00} used three different methods to estimate the spin
temperature. As our aim is to compare ${\cal M}_s$ with values
derived from the LOS integrated HI emission profiles,
we use their LOS averaged spin temperature (or $T_{\rm ew}$).
The median value of the spin temperature is 43 K.

\begin{figure*}[tbh]
\centering
\includegraphics[scale=.47]{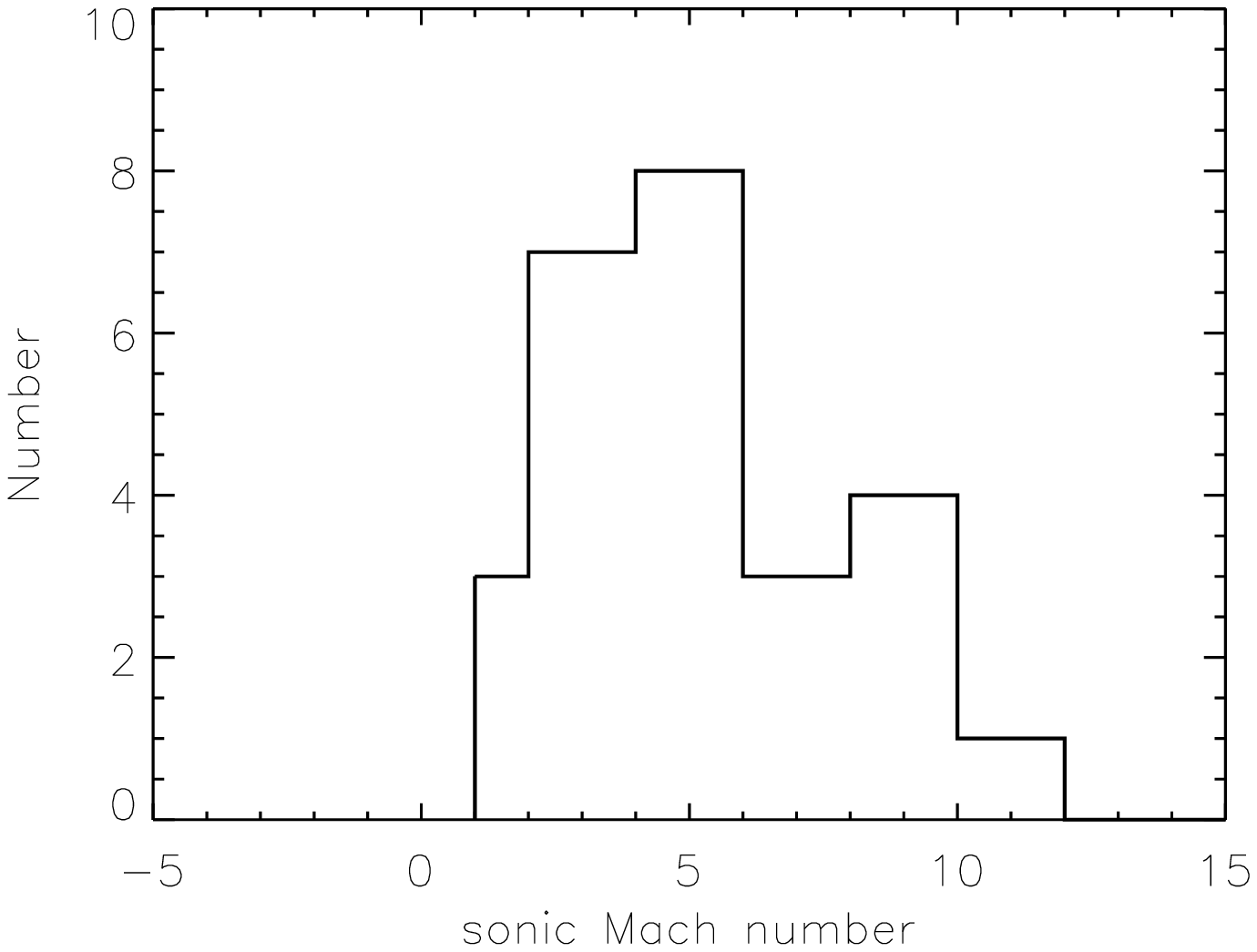}
\includegraphics[scale=.5]{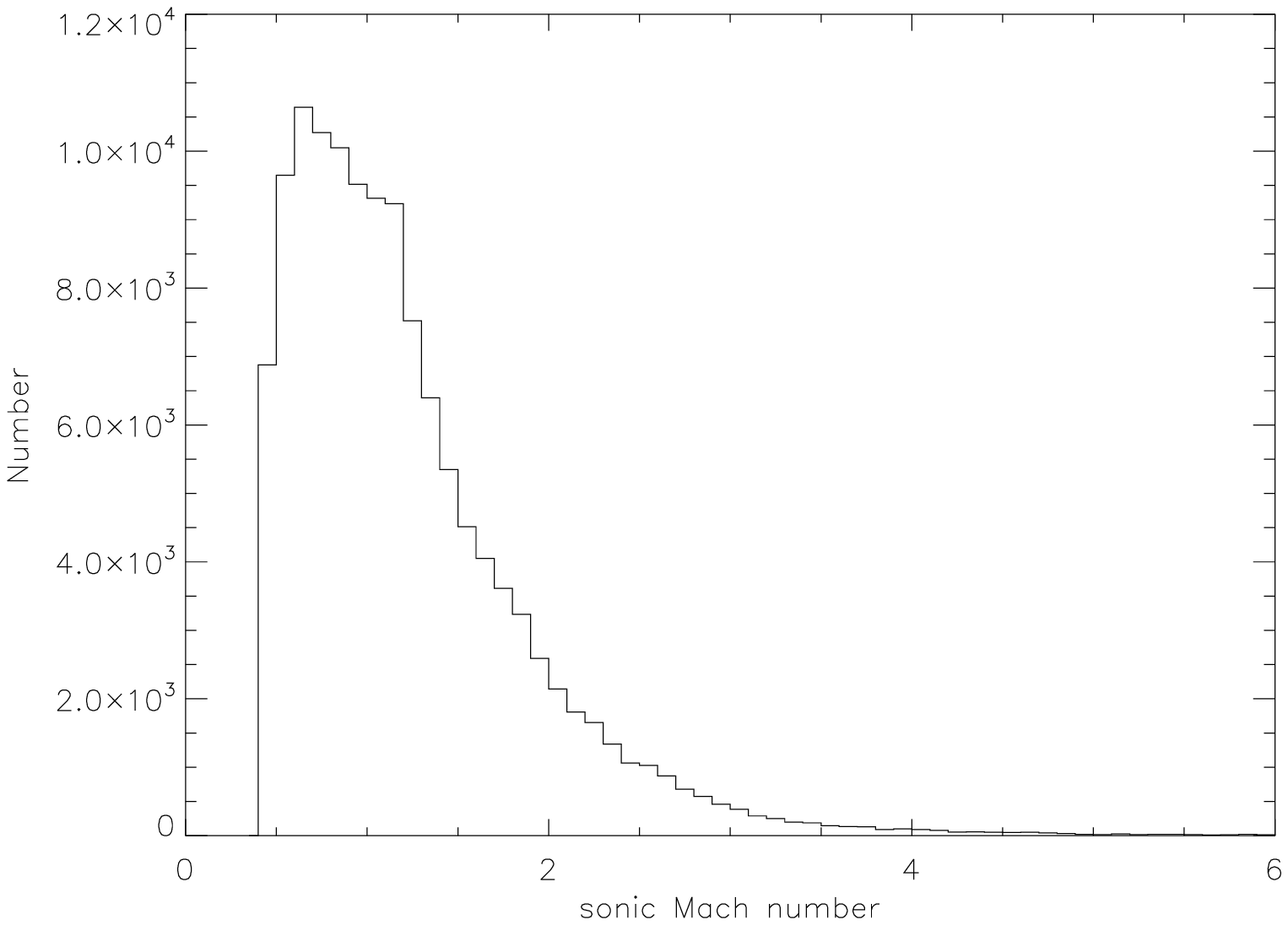}
\caption{(Left) Histogram of the sonic Mach number of the CNM clouds in the SMC derived 
from the ratio $T_s/T_{k,max}$ using the HI absorption profiles 
against radio continuum sources. 
(Right) Histogram of the sonic Mach number from the ${\cal M}_s$ map derived 
using the higher statistical moments (see Figure~\ref{fig:ms_var}).  
}
\label{fig:hist}
\end{figure*}

A histogram of derived ${\cal M}_s$ values is shown  in Figure~\ref{fig:hist}. 
The median Mach number for the whole sample of 29 sources is ${\cal M}_s=4.7$,
and the histogram peaks around ${\cal M}_s=3.5-4$.
This suggests that the internal CNM macroscopic motions are highly supersonic.
To compare these values with the ones derived using the method of 
higher-order statistical moments we show a
histogram of data points from the ${\cal M}_s$ map in the right panel of Figure~\ref{fig:hist}.
While the observed histogram suffers from the low number statistics, due to a small
number of suitable (strong) sources,
the two histograms have a similar shape: almost a Gaussian central distribution
with a tail at higher Mach values.
Obviously, the observationally inferred ${\cal M}_s$ values suggest at least a factor of 2 higher
level of turbulence across the SMC than what we derived using higher statistical moments.
This is not surprising as the observed values trace predominately the CNM, while
the ${\cal M}_s$ map was derived using the HI column density image and therefore
is affected by both CNM and WNM.
%Thus it would be surprising to see drastic similarities of the sonic Mach 
%number of both the CNM and the WNM.
%This could happen if the CNM is formed out of the turbulent WNM
%and therefore carries the signature of the same turbulent spectrum.
We discuss this further in \S~ \ref{sec:Discussion}. 
%It is likely that that the
%skewness and kurtosis of the HI column density are biased towards cold gas and therefore
%trace the CNM turbulence.
In any case, to sample better CNM turbulence across the SMC 
future sensitive HI absorption observations  of continuum sources
behind the SMC are clearly needed.

\section{Spatial Power Spectrum of Column Density}

\label{ps}

The slope of the Fourier transform of the two point autocorrelation 
function, or the spatial power spectrum, also provides information on important 
properties of turbulence flows. In particular,
the slope of the power spectrum is known to depend on the sonic Mach number.
With a database of numerical simulations 
we explore whether the slope of the spatial power spectrum depends 
also on the Alfv\'en Mach number. 
The Alfv\'en Mach number dependence would be especially 
interesting as we could use higher-order
statistical moments to estimate the sonic Mach number, and from there 
the Alfv\'en Mach number and the 
strength of the magnetic field.

The power spectrum is defined as:
\begin{equation}
P(k)=\sum_{|\vec{k}=k|}{A(\vec{k})\cdot{A}^{*}(\vec{k}})
\end{equation} 
where $k$ is the wavenumber  and $A(\vec{k})$ is the Fourier transform.
Stanimirovic \& Lazarian (2001) estimated the power-law slope of $-3.3$
for the spatial power spectrum of the HI column density image of the SMC.
%We note that in the power spectrum
%of the SMC, however, there are several bumps at 
%the smaller scales, which may indicate the end of the 
%intertial range.  Therefore we only look at the 
%slope for the SMC over scales k=4.4kpc to 100pc, which 
%is $\approx -2.0$.
We measure the power spectrum slope for various simulated column densities.

\begin{figure*}[tbh]
\centering
\includegraphics[scale=.5]{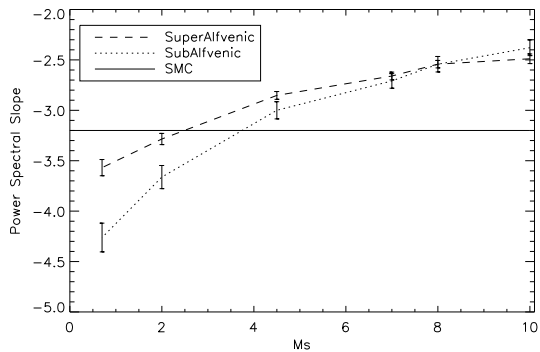}
\caption{Power spectral slope vs. sonic Mach number for a sub-Alfv\'enic (${\cal M}_{A}=2.0$, dotted
line) and a  super-Alfv\'enic (${\cal M}_{A}=0.7$, dashed line) 
simulation. The spectral slope of the SMC ($-3.3$) is shown as a straight line.}
\label{fig:power}
\end{figure*}

Figure ~\ref{fig:power} shows how the power spectral slope changes with the 
sonic Mach number.  The slope is increasingly shallow for supersonic models and levels
off for very high Mach number turbulence. This is expected as higher Mach number
turbulence has more density irregularities and more power on small scales.  
We have shown in this figure slopes for two values of ${\cal M}_{A}$: 0.7 (dashed line)
and 2.0 (dotted line). For both strong and weak magnetic fields, the power spectrum
slope increases with ${\cal M}_{s}$, however this increase is steeper for
super-Alfvenic and subsonic turbulence.  
For all cases the error bars were calculated from the standard
deviation in the power spectral slope derived from column density images 
for three different LOS orientations.

The SMC has a power spectrum slope of $-3.3$ and spatially
mostly ${\cal M}_{s} <2$.
Based on the difference between slopes expected for sub-Alfv\'enic and super-Alfv\'enic turbulence in 
Figure~\ref{fig:power}, it is likely that the super-Alfv\'enic description fits better the
slope of the SMC. 
%which suggests supersonic values (around ${\cal M}_{s} \sim 3$).  
%This is consistant with the analysis of moments 
%and $\frac{T_k}{T_s}$ done in the previous sections.

\section{Bispectrum}
\label{sec:bispectrum}
\subsection{Calculation of the bispectrum}

The bispectrum is closely related to the power spectrum. 
In a discrete system, the power spectrum is defined by Equation (11).
In a similar way, the bispectrum is defined as:
\begin{equation}
B(k_{1},k_{2})=\sum_{\vec{k_{1}}=k_{1}}\sum_{\vec{k_{2}}=k_{2}} 
A(\vec{k_{1}}) \cdot A(\vec{k_{2}}) \cdot {A}^{*}(\vec{k_{1}}+\vec{k_{2}})
\label{eq:bispectra}
\end{equation}
\noindent
where $\vec{k_{1}}$ and $\vec{k_{2}}$ are the wave numbers of two 
interacting waves, and $A(\vec{k})$ 
is the original discrete time series data with finite number of elements
with $A^{*}(\vec{k})$ 
representing the complex conjugate of $A(\vec{k})$.

\begin{figure*}[htb]
\centering
\includegraphics[scale=.8]{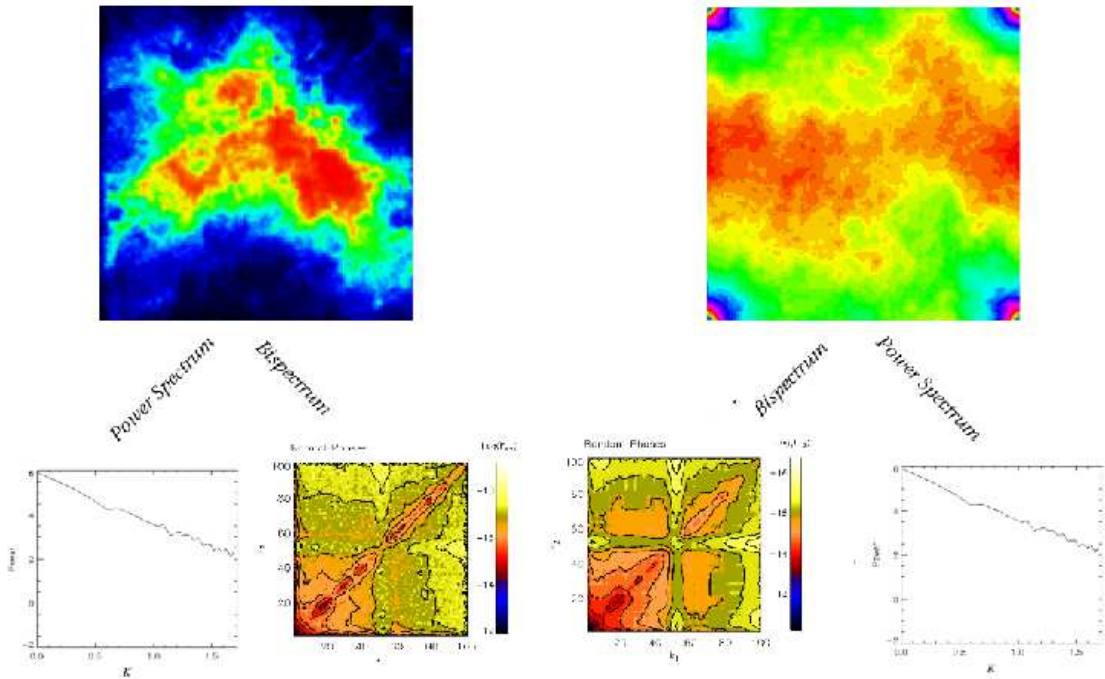} 
\caption{ This figure illustrates differences between the bispectrum and 
the spatial power spectrum. 
The image at the top left is the original SMC HI column density image, 
while the image on the right has the same amplitudes however the 
phases have been randomized with a random Gaussian distribution. 
As expected, the power spectrum of the 
two images is identical, but the bispectrum shows a significant difference.
\label{fig:example}}
\end{figure*}

The bispectrum is a complex function which measures both phase and magnitude 
information between different wave modes. 
As this is the first application of the bispectrum on the HI data
we show in Figure~\ref{fig:example} a visual example of the difference 
between the power spectrum and the bispectrum. This figure shows the 
original HI column density image of the SMC (top left), and the same image 
but with manipulated phases (top right).  
To obtain the top right image we Fourier 
transformed the SMC image and randomized the phases with a Gaussian random distribution 
but left the \textit{amplitudes} the same.  
Even though the phases of the top images are very different, 
the power spectrum uses only amplitude
information and is identical for both images,
as shown in the bottom panels of Figure ~\ref{fig:example}.
However, the bispectrum looks very different for the two
images, as expected, and offers an insight into the
phase information.

In practice, our calculation of the bispectrum involves performing a Fast 
Fourier Transform (FFT) of the column density functions and the application of 
Equation~\ref{eq:bispectra}. 
We randomly choose wavevectors and their directions, $k_{1}$ and $k_{2}$ 
and iterate over them, calculating $k_{3}=k_{1}+k_{2}$.
 We limit the maximum length of the wave vectors to 
half of the box size. We normalize direction vectors to unity, calculate 
positions in Fourier space, and finally, compute the bispectrum, which 
yields a complex 2D array.  
When plotting the bispectrum (Figures 12-15) we plot bispectral 
amplitudes, which  give information 
about the degree of mode correlations in the original column density distribution.

\subsection{Bispectrum of MHD Simulations}

\begin{figure*}[htb]
\centering
\includegraphics[scale=.5]{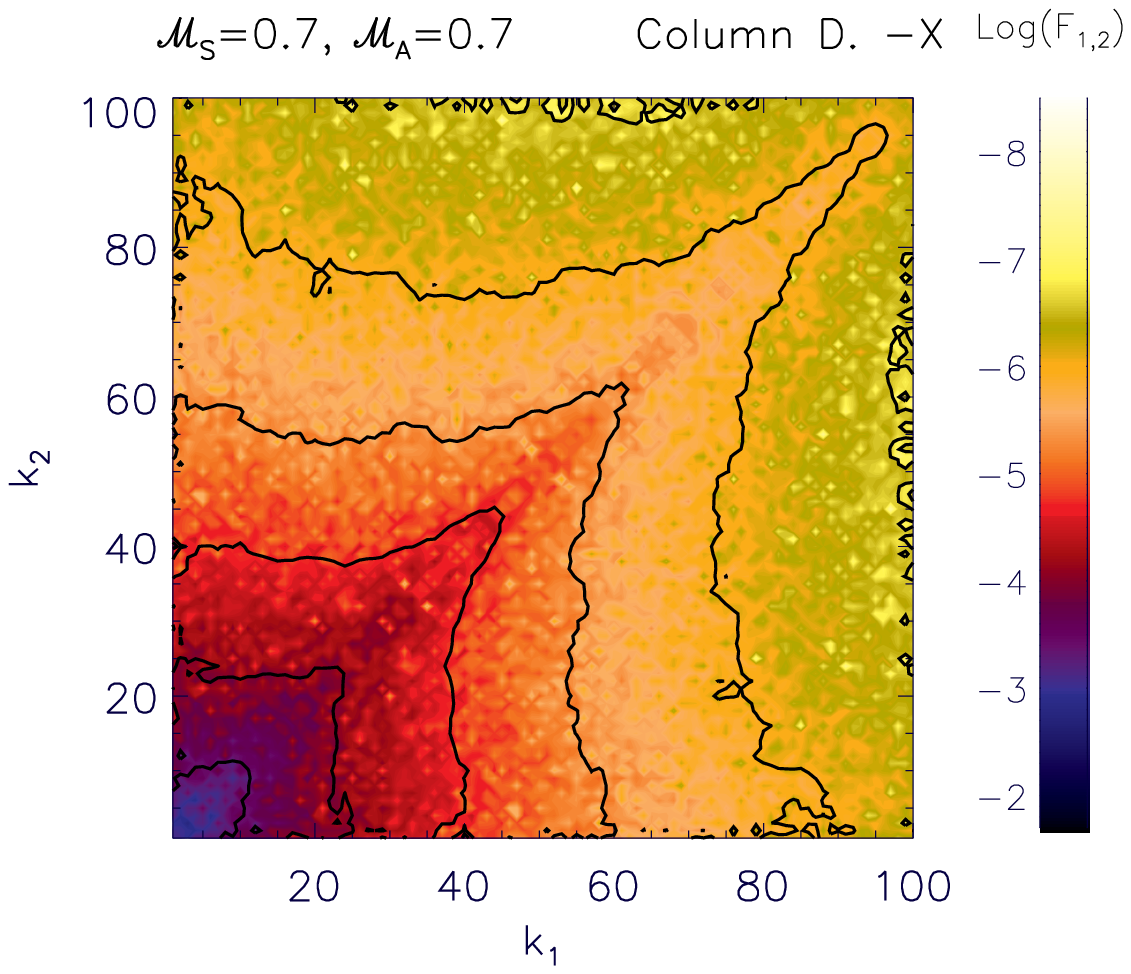} 
\includegraphics[scale=.5]{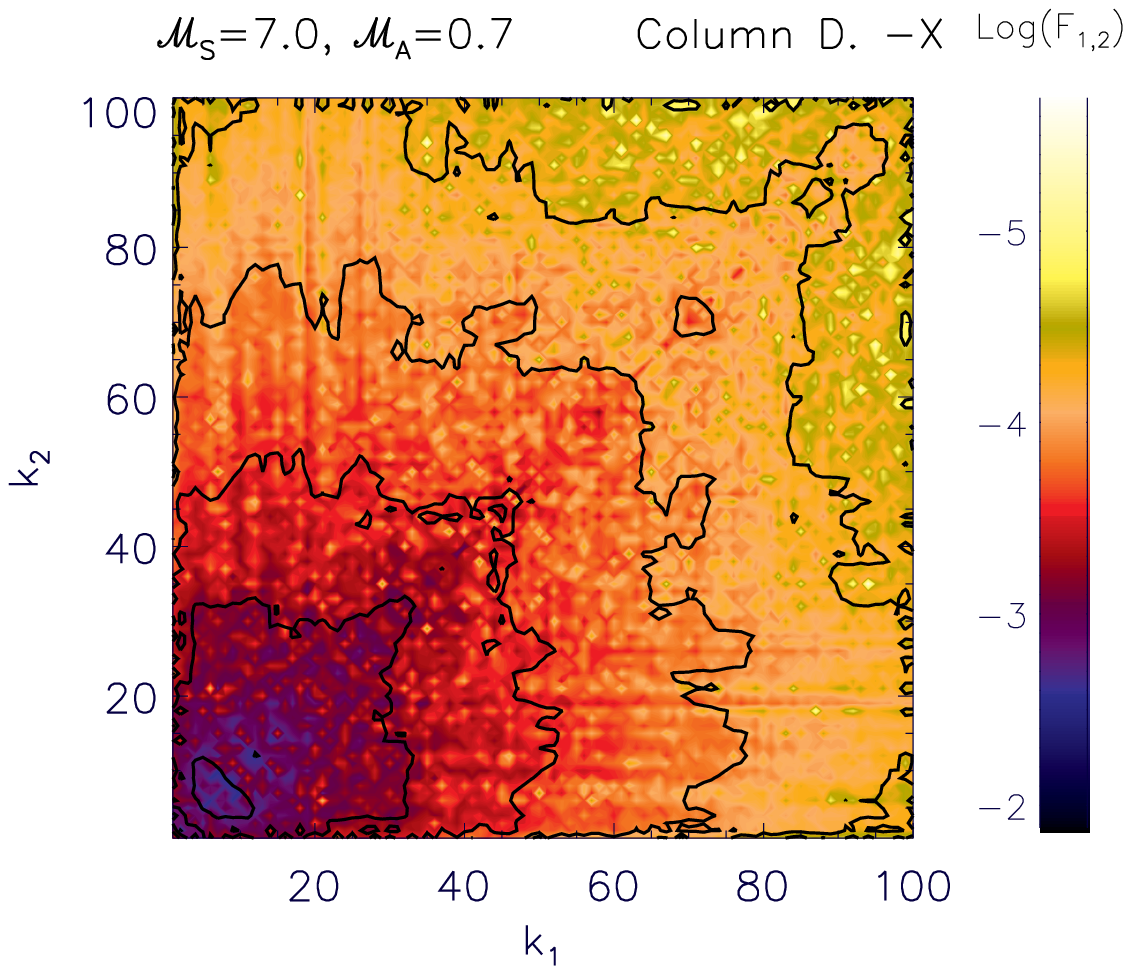} 
\caption{The amplitude of the bispectrum for the scaled simulated
column density, derived by integrating along the x direction. 
The left plot shows a subsonic model, while the right plot is for a supersonic model.  
Both models have ${\cal M}_A$=0.7.  These figures show the degree of correlation 
between wavenumbers $k_{1}$ and $k_{2}$. 
The supersonic model has higher bispectral amplitudes, and more
circular isocontours, therefore a stronger correlation between wave modes.
\label{fig:bispsimms}}
\end{figure*}

\begin{figure*}[htb]
\centering
\includegraphics[scale=.5]{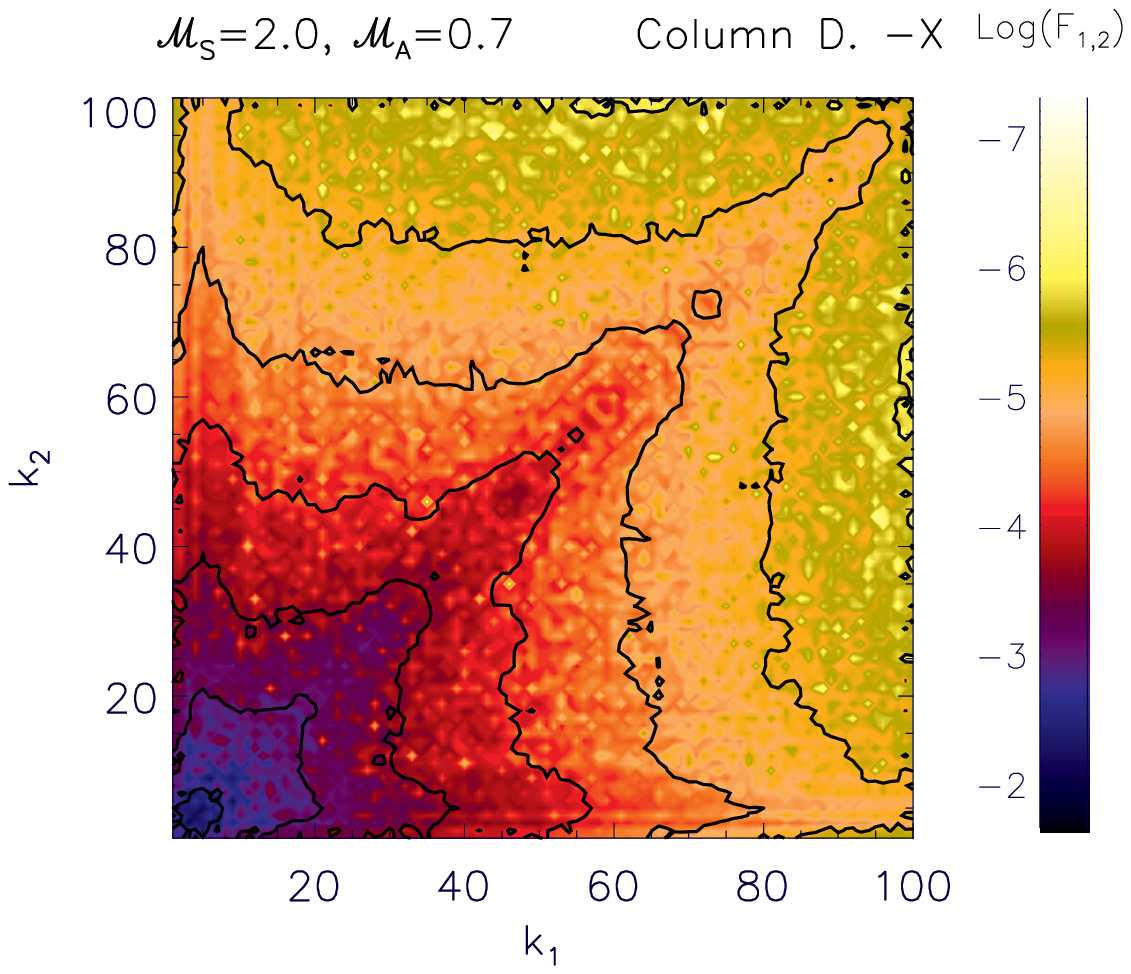} 
\includegraphics[scale=.5]{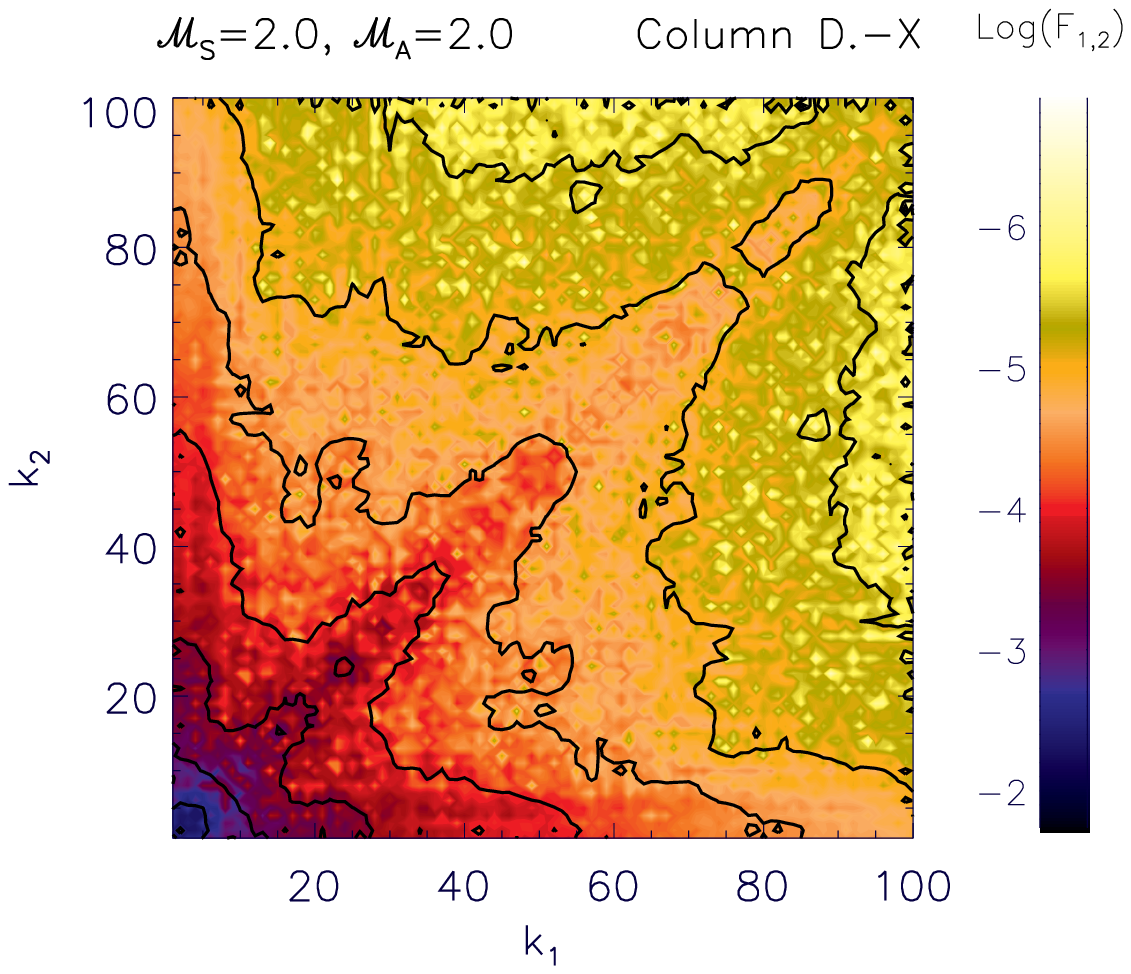} 
\caption{ The amplitude of  the bispectrum for the scaled  
simulated column density. 
The left plot shows a sub-Alfv\'enic model, while the right 
plot is for a super-Alfv\'enic model.  Both models have ${\cal M}_s$=2.0.  
These figures show the degree of correlation between wavenumbers $k_{1}$ 
and $k_{2}$.
While the difference in amplitudes is less pronounced than in the previous figure, 
the model with a higher magnetic field (i.e. sub-Alfv\'enic)
has more circular isocontours and therefore a slightly
enhanced correlations between wave modes.
\label{fig:bispsimma}}
\end{figure*}

In order to interpret the bispectrum of the SMC HI column density 
we first compute the bispectrum of the  MHD simulations (scaled by the standard 
score method and without applying cloud boundaries) and 
look for dependence of wave-wave correlations on
the sonic and Alfv\'en Mach numbers.
Figure~\ref{fig:bispsimms} shows the bispectrum of simulated
column density distributions for the case of fixed
${\cal M}_{A}=0.7$ and two extreme cases of ${\cal M}_{s}$: 0.7 and 7.0.
In Figure~\ref{fig:bispsimma} we fix ${\cal M}_{s}=2.0$ and look
at two cases of ${\cal M}_{A}$: 0.7 and 2.0.

%As expected, the scaled 
%models look similar to those shown in BFKL, only the amplitudes have 
%changed due to the scaling of original column density distributions.

The first thing to notice in the bispectral contour maps is that the $k_1=k_2$
case always shows high correlation since this is a trivial case of two
wave numbers being the same. For $k_1 \neq k_2$ the bispectral amplitude
and isocontour shape vary with turbulent properties but generally
amplitude decreases gradually radially from $k_1=k_2=0$.  
Models with circularly/broad-shaped isocontours
have high amplitudes and wave-wave correlations, while models with more narrow 
isocontours have lower amplitudes and therefore weaker correlations.
For a fixed ${\cal M}_{A}$, supersonic models show a higher degree 
of wave-wave correlations over subsonic models.
For a fixed ${\cal M}_{s}$, models with a higher magnetic
field (sub-Alfv\'enic, e.g. ${\cal M}_{A}=$0.7) show 
somewhat stronger correlations then the models with
a weaker magnetic field (super-Alfv\'enic) although this difference is
not striking.

\subsection{Bispectrum of the SMC}

\begin{figure*}[htb]
\centering
\includegraphics[scale=.8]{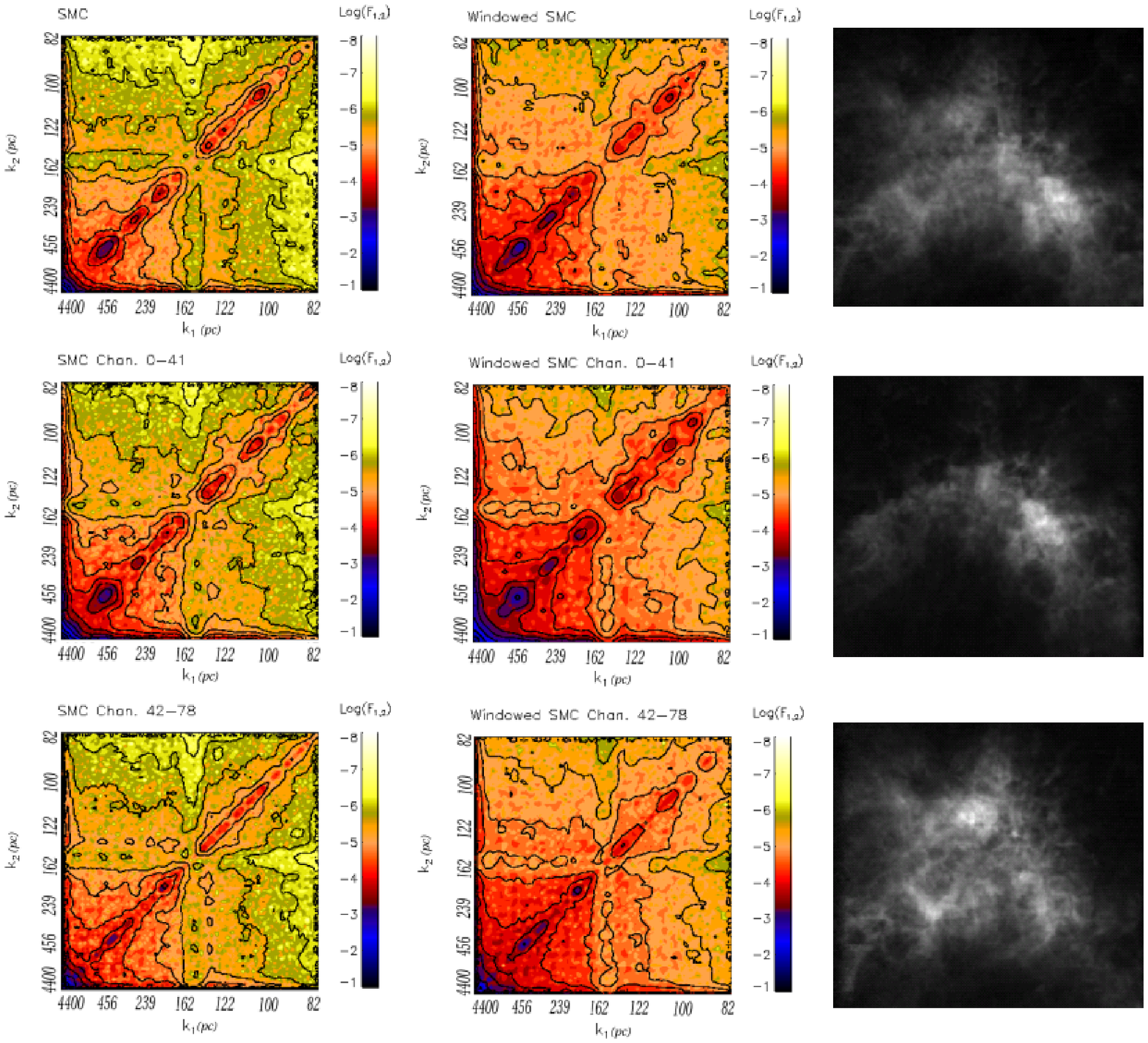} 
\caption{The amplitude of the bispectrum for the SMC HI column density. 
The left column shows the bispectrum for the scaled unwindowed SMC image, while the 
right column shows the effects of a Hanning window on the bispectrum.  
The first row shows the SMC column density image derived by 
integrating over all 78 velocity channels. 
The second row is for the image derived by integrating
over channels 1-42, while the third row shows the 
bispectrum of the image obtained by integrating the HI data cube over
channels 43-78. The appropriate column density images used to derive the bispectrum
are shown in the right-hand column.
The bispectrum shows the degree of correlation between 
wavenumbers $k_{1}$ and $k_{2}$ in at 
varying depths of integration. 
We display the bispectral amplitude for spatial scales 4400--80 pc as the bispectrum
is very noisy at smaller scales.
  \label{fig:bispectraglobal}}
\end{figure*}

The bispectrum of the SMC HI column density is shown in Figure~\ref{fig:bispectraglobal}. 
The top row (first column) shows the bispectrum of the column density derived by integrating 
all 78 velocity channels. The middle and bottom  rows show the bispectrum of 
the first (for channels 1 to 42) 
and the last half (for channels 43 to 78) of the data cube. 
To facilitate interpretation the wavenumbers $ k_1$ and $k_2$ are shown in terms of linear 
size\footnote{The linear scale $L=4.4$ kpc corresponds to the largest angular scale
covered in the original image used to calculate the bispectrum, $=576*30''=15,360''$.
We scale then all wavenumbers $k$ by $2\times15,360''/k$ to express wavenumbers in terms of physical
linear size at the distance of the SMC (60 kpc).}.
While there are small differences in the bispectral amplitudes, there is
essentially no significant difference in the contour shape for the three bispectra.

To properly compare the SMC bispectrum with simulations
we apply a \textit{windowing} function on the SMC column density image
to simulate the effect of periodic boundaries.
We use the Hanning window $w(n)$ function, defined as: 
\begin{equation}
w(n)=0.5 \left (1- \cos \frac{2\pi n}{N-1} \right)
\end{equation}
where $N$ is the number of pixels along the x or y axis, and $n$ ranges from 0 to $N-1$. 
This windowing function makes the map periodic in the Fourier space.
The second column of Figure~\ref{fig:bispectraglobal} shows the resultant
bispectra for all three cases of HI column density integrations.
The windowing seems to slightly 
change bispectral amplitudes on the large 
scales, but leaves the general structure of the bispectral contours intact, 
although some smoothing effects are observed. 
The increase in large scale correlations is at a level of $\sim10$\%.

\begin{figure*}[htb]
\centering
\includegraphics[scale=.5]{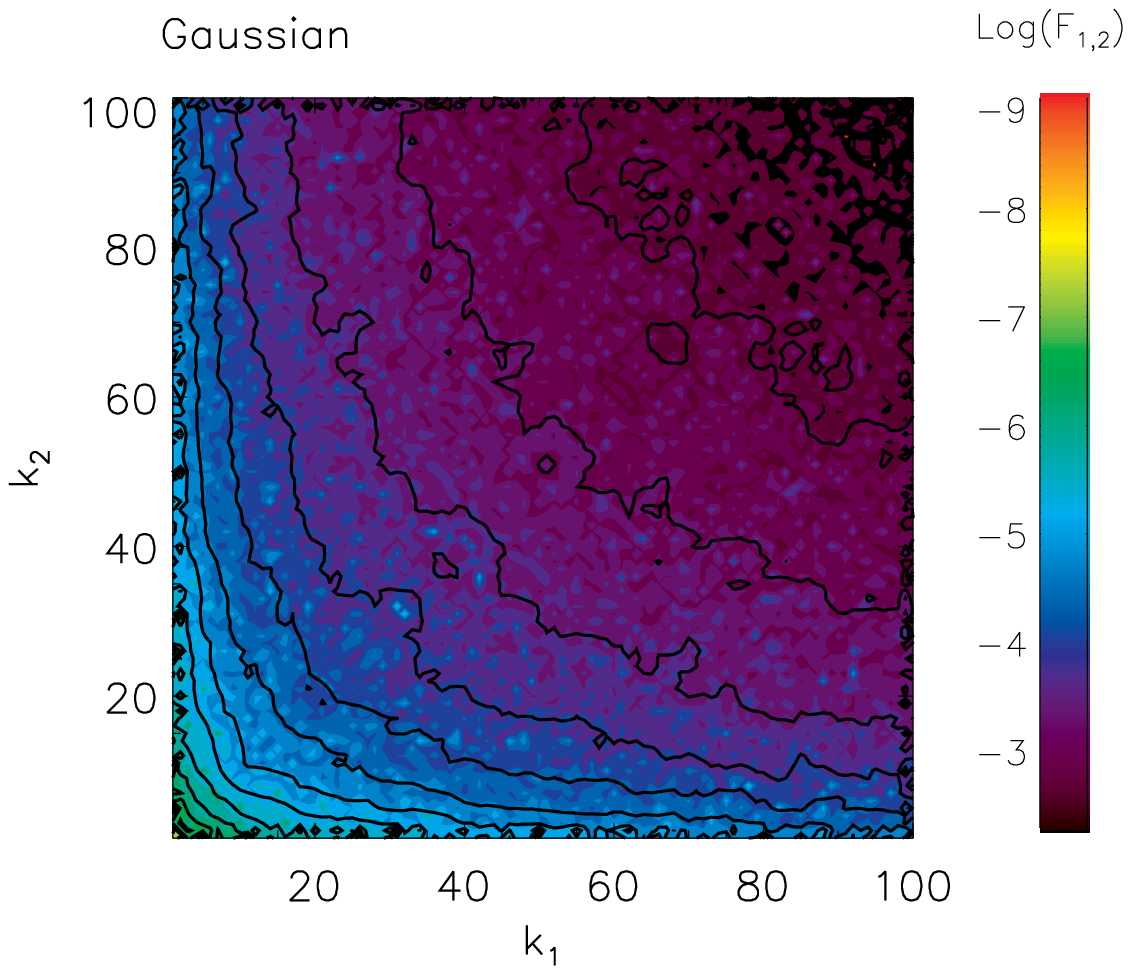}
\includegraphics[scale=.5]{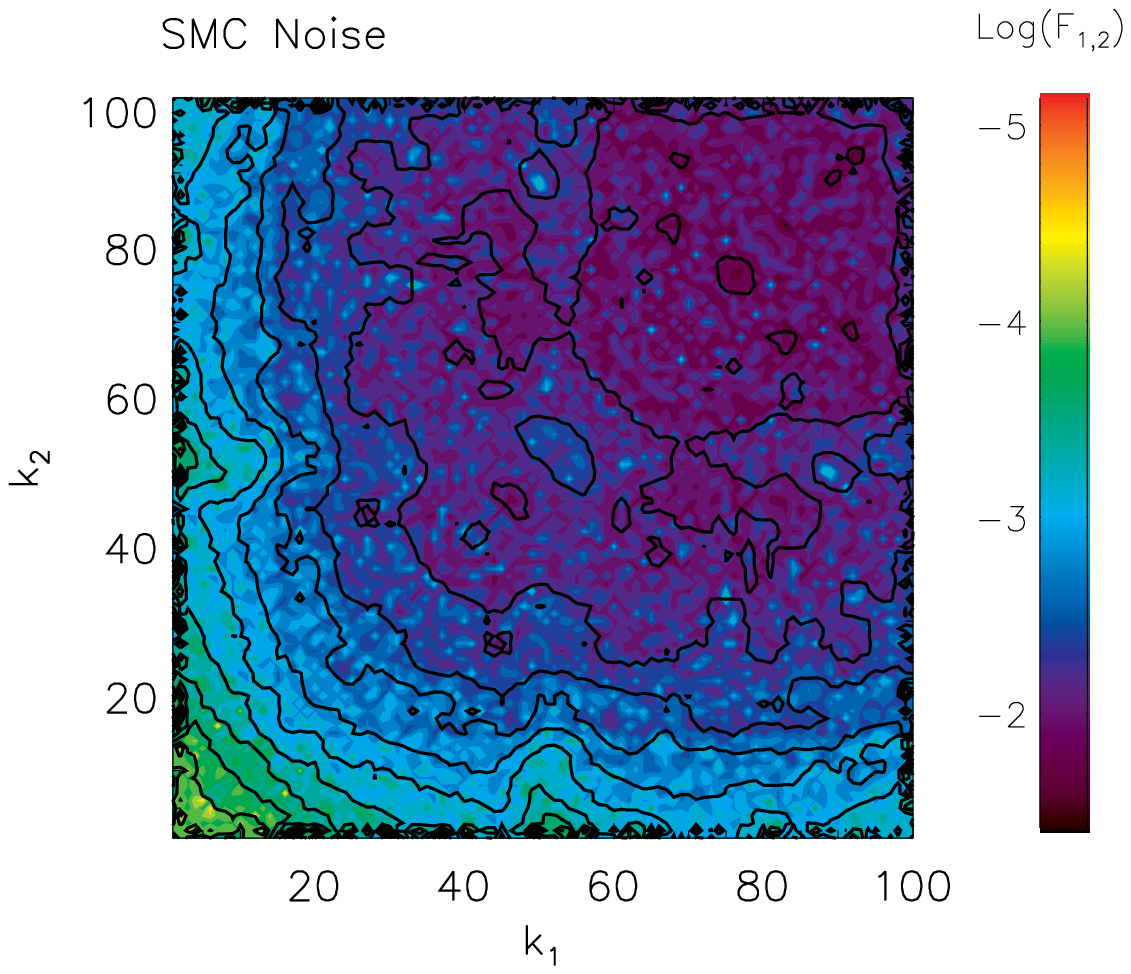}
\caption{(Left) The amplitude of the bispectrum of a pure
Gaussian distribution. It gradually increases from almost zero at small $k$
to $10^{-3}$ at large $k$.
(Right) The amplitude of the bispectrum derived for a single emission-less
channel from the SMC HI data cube.
\label{fig:noise}}
\end{figure*}

Before interpreting the SMC bispectrum
we briefly investigate how the presence of noise in the observational data affects 
the bispectrum.
We made the noise image of the HI column density by taking a single 
velocity channel from the SMC HI data cube without
HI emission and integrating this image over the whole velocity range.
The bispectrum of the noise image (normalized using the standard score method) 
is shown in Figure ~\ref{fig:noise} (right).
In the case of purely Gaussian noise, we expect a random distribution over the whole
space of  wavelengths. As a result, the bispectrum
will have a gradual increase in amplitude, with the highest
amplitude being at large $k$ (or small scale).
Figure ~\ref{fig:noise} (left) illustrates this effect for a
pure Gaussian distribution.
The bispectrum of the SMC noise image shows essentially the same trend,
suggesting that we are dealing mainly with Gaussian (random) noise.
As is obvious from Figures 12-14, the bispectral amplitude increases
in the opposite direction (from large to small scales).
The effect of noise is therefore the most important at the smallest spatial scales,
and we keep this in mind when interpreting the bispectrum.

%INTERPRETATION:
Figure~\ref{fig:bispectraglobal} shows the degree of wave-wave 
correlations in the HI distribution.
Modes are obviously strongly correlated at the largest scales and 
show weaker correlation at small scale. This trend is similar to what
was found for various MHD simulations, and obviously very different
from the case of Gaussian noise. 
In addition, the $ k_1=k_2$ line shows the strongest correlations, 
as is also seen in the simulations.
However, contrary to simulations where the level of correlation
gradually decreases along this line, we see significant
small-scale variations. Several local peaks are noticeable and
a strong break at mid-scales.
The break occurs around $k_1=k_2 \approx 160$ pc 
and is seen in all three integrations of the HI data cube.
This could be due to a lack of interactions between turbulent eddies
at the scale of $\sim160$ pc, caused possibly by
numerous expanding shells in the SMC.
Hatzidimitriou et al. (2005) and Stanimirovic (2007) showed that shells 
in the SMC range in size from 30 to 800 kpc, however the radius distribution peaks at  
$\sim60$ pc. The break in the bispectrum may be signifying the lack of correlations
on scales close to the typical shell diameter, or some additional physical
processes.

We can also compare the bispectrum of the SMC with the bispectra of 
various simulations.
By visual inspection of Figures 12-13, the closest simulation to 
the SMC in terms of contour shapes is the one for ${\cal M}_s= 2.0 $ and
${\cal M}_A= 2.0 $. Further work in this area is required to 
derive a real measure to quantitatively compare observed and simulated bispectra.

\section{Discussion and Summary}

\label{sec:Discussion}

\subsection{Turbulence properties of the HI gas in the SMC}

Using 3rd and 4th statistical moments of the HI column density image
and boothstreping turbulent information from a database of
isothermal MHD simulations, we have mapped spatial variations
of the sonic Mach number across the SMC. While most of the HI seen in emission 
in the SMC appears to be subsonic or transonic, several supersonic
regions have emerged from our study.  It is interesting that these
regions do not correlate well with the most recent sites
of star formation and seem to point out to large scale shearing or tidal flows.
Commonly, it is believed that supernovae and superbubbles
are the main drivers of galactic turbulence (McCray \& Snow 1979), with
a typical size of $\sim100$ pc. While we do not have high enough resolution
to see changes on such small scales ($\sim10'$) in our derived map of ${\cal M}_s$,
most of the star-forming bar of the SMC appears to have subsonic or transonic properties 
when viewed at resolution of 30$'$.

The most turbulent regions in the SMC may be tracing some kind
of shearing flows between the SMC bar and the surrounding HI. This suggests that 
SMC's chaotic history with the LMC  and our own Milky Way has probably left
strong turbulent imprints on the HI gas.
The lack of a turnover in the HI spatial power spectrum on the largest
observed scales is also indicative of the fact that turbulent energy injection
happens largely on scales larger than the size of the SMC  \cite[]{Stan01}.
Similarly, Goldman (2000) suggested that the HI turbulence in the SMC
was induced by large-scale flows from tidal interactions
with the Milky Way and the LMC about $2\times10^8$ yrs ago. 
Such large-scale bulk flows could have 
generated turbulence through shear instabilities, and
this turbulence has not have had enough time to decay.

Most of the HI in the SMC has a sonic Mach number of 1-2. This
is on average at least two times smaller than what we inferred from
HI absorption observations for the CNM in the SMC, ${\cal M}_s\sim$3.5-4.
Similarly, for the CNM in the Milky Way Heiles \& Troland (2003) 
found ${\cal M}_s \sim3$ with a large dispersion.
A sonic Mach number of about 4-5 is commonly assumed for cold gas in molecular
clouds (Federrath et al. 2009). For example, Heyer et al. (2006) measured from CO observations
${\cal M}_s=4.2\pm0.17$ for the Rosette molecular cloud,
and ${\cal M}_s=4.7\pm0.12$ for G216-2.5.
On the other hand, Hill et al. (2008) found that the distribution of 
the warm ionized medium (WIM) in the 
Milky Way can be best fit by models for 
mildly supersonic turbulence with ${\cal M}_s\sim1.4-2.4$.
Turbulent properties of the HI in emission in the SMC are therefore closer to
properties of the WIM in the Milky Way than properties of the CNM. 
This may suggest a large fraction of 
warm relative to cold HI being traced in HI emission.

In the Milky Way, the HI is known to consist of at least two components 
with different temperature: the WNM with $T_{warm}=6000$~K and 
the CNM with $T_{cold}=70$~K. In addition, there could be a substantial amount of 
gas at intermediate temperatures \cite[]{Heiles03}. 
Due to its lower metallicity, the HI in the SMC has different properties.
Dickey et al. (2000) found $T_{cold}=40$~K, in agreement with theoretical expectations
by Wolfire et al. (1995) whereby the existence of the two-phase
medium is possible only at higher
pressures compared with the range that applies for
solar neighborhood conditions. They also estimated the fraction of cold HI in the
SMC to be $\la15$\%. This is lower than $\sim25$\% found for the Milky Way.

\begin{figure*}[tbh]\centering
\includegraphics[scale=.7]{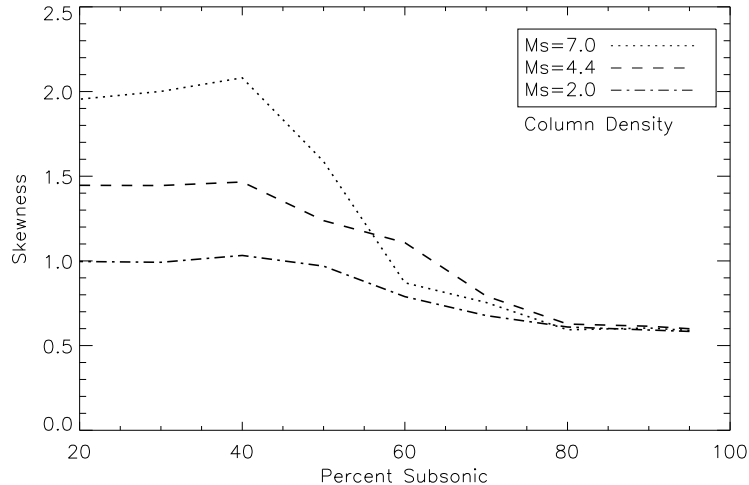}
\caption{Skewness vs. percent of gas that is subsonic 
for column density of a $512^3$cube weighted with 
supersonic and subsonic gas.  Supersonic gas generally dominates 
the skewness of the column density PDF.}
\label{fig:ex}
\end{figure*}

As our simulations are isothermal it is obvious to wonder 
how does the multi-phase HI affect our statistics and conclusions.
We investigate this in Figure~\ref{fig:ex} by producing a simulated data cube
from a weighted combination of two cubes, one subsonic (${\cal M}_{s}$=0.7) and 
one supersonic.
The subsonic cube represents contribution from warmer gas, while the supersonic
cube represents the cold gas. 
We combined the two cubes with different emphasis on warm 
vs cold gas, obtained the column density image of the resultant cube, and 
calculated its moments.

Figure~\ref{fig:ex} shows that 
for the case when the supersonic cube has ${\cal M}_{s}$=2.0 
skewness of the final cube with up to 50\% of subsonic gas
will be biased towards supersonic gas and will appear dominated 
by cold gas. If we increase the sonic Mach number of
the supersonic cube to 4.0, the dominance of the higher turbulence is even more
pronounced. A cube with up to 60-70\% of subsonic gas and 25\% of supersonic gas,
will still have high skewness biased by the supersonic contribution.
Considering that the HI column density image
results in the mean ${\cal M}_{s}=1-2$, significantly lower than what is expected for 
the cold HI, the fraction of the CNM along any LOS is most likely  
$\la25$\%. This supports the Dickey et al. (2000) estimate of 
the fraction of cold HI in the SMC being about 15\%.

\subsection{Is the HI in the SMC sub-Alfv\'enic or Super-Alfv\'enic?}

Two different statistical approaches in our study suggest that 
the HI gas in the SMC seen in emission is super-Alfv\'enic.
As we have shown in Figure ~\ref{fig:power}, in addition to the sonic Mach number
the spectral slope of the spatial power spectrum 
is sensitive to the Alf\'venic Mach number for ${\cal M}_{s}<2$. 
The sub-Alfv\'enic models generally show steeper slopes due to large scale influence of 
magnetic fields.
Thus, if one independently knows the sonic Mach number, it is possible to estimate 
the Alf\'venic one using just the column density data.
While the dependence of the spectral slope on the Alf\'venic Mach number 
has not received much attention in the past,
it is somewhat expected. Essentially, magnetization decreases compression in the shocks. 
Strong magnetic forces mix up density clumps preventing formation of 
isolated peaks, which results in a steeper spectrum.
In addition, in the sub-Alfv\'enic case we expect
oblique shocks to be disrupted by Alfv\'en shearing, 
which in turn, produces more small scale shocks
\cite[]{beresnyak05}. 

Another indication that the HI in the SMC is super-Alfv\'enic comes from the bispectrum.
The very sharp decrease in the bispectral amplitudes from large to small scales observed
for the SMC is the closest to the trend found for simulated data for
the case of ${\cal M}_{s}\sim2$ and  ${\cal M}_{A}\sim2$. 
Detailed comparison between simulated and observed bispectra awaits future work,
however this qualitative comparison is certainly encouraging.

Assuming on average ${\cal M}_{s}\sim1-2$, the power 
spectrum slope suggests a super-Alfv\'enic HI in the SMC with ${\cal M}_{A}\sim1-3$.
This is generally in agreement with the observationally inferred 
strength of the magnetic field by Mao et al. (2008). Using
their estimate for $B_{\rm ext}=2$ $\mu$G, a radius of the SMC of 2 kpc, the total hydrogen mass
of $4.2\times10^{9}$ M$_{\odot}$, and a typical velocity dispersion of 20 \kms 
(Stanimirovic et al. 2004), we estimate ${\cal M}_{A}\sim3$.
As the Alf\'venic Mach number shows the nature of the interplay
between gas pressure and magnetic fields, it appears that the gas pressure in the SMC
dominates over the magnetic pressure.

\subsection{Intermittency in mode correlations? }

Our bispectrum analysis of the SMC HI data was the first attempt to apply bispectrum on
observed astrophysical data. While more detailed comparison
between observations and simulations awaits future work, 
we clearly see trends in the bispectral amplitudes similar
to what was found for simulations of supersonic MHD turbulence.
The most interesting finding is, however, the effect of small-scale variations
in the $k_1=k_2$ correlations and a strong break in correlations at a scale of 160 pc.
Such small-scale variations, or jumps, have not been seen in the bispectrum
of simulated data. We can speculate about several possible scenarios that 
could explain their existence.
The jumps could be caused by the energy injection due to processes other than turbulence
affecting specific spatial scales. Alternatively, the jumps may be marking
the presence of colder or multi-phase gas. 
Similarly, the observed break in the bispectrum at about 160 pc is intriguing.
As we already pointed out, it is interesting that most expanding shells in the SMC (more
than 500 were cataloged so far) have a diameter of $\sim120$ pc.
The break could be due to the lack of correlations on scales similar to 
the distance between two shell centers.
Obviously this will require further studies.

\subsection{Limitations of the present study}

A natural question to ask is how results presented in this paper 
depend on the resolution of numerical simulations. 
For example, Kritsuk et al. (2007) investigated how
resolution of numerical simulations affects the power spectrum of 
density.  These authors found that 
the spectral index estimates based on low resolution simulations bear 
large uncertainties due to the bottle neck contamination, and 
that the power spectra of $512^3$ simulations are substantially 
shallower then  models with resolution of $1024^3$. 
However, while Kritsuk et al. (2007) only examined hydrodynamic turbulence, 
Beresnyak et al. (2008) showed that the slopes were very different 
between the MHD and pure hydrodynamic cases. For instance, the slopes
 for hydrodynamic simulations showed a pronounced and well defined 
bottleneck effect, while the
 MHD slopes were much less affected. This is indicative of 
MHD turbulence being less local than the Kolmogorov turbulence,
and suggests that our simulations will be less affected by resolution.
In addition, Kritsuk et al. (2007) found a
difference in the slope between hydrodynamic $512^3$ and $1024^3$
simulations to be 0.17. This would result in a change of $d{\cal M}_{s}\sim0.5$
only and will not change our interpretation.
We also add that in the case of higher statistical moments and
the bispectrum BFKL confirmed trends noticed by KLB at lower resolution of
128$^3$.

Another issue that should be further addressed and that could affect our results 
is the type of numerical forcing of turbulence. 
Federrath et al. (2009) recently investigated the effects of solenoidal 
vs. compressive (divergence-free vs. curl-free) forcing on a variety of statistics 
including PDFs and higher order moments. They found that both types of driving mechanisms 
are compatible with observations of molecular clouds 
however, depending on the data studied, one type could 
be superior then the other in terms of the statistics and reproduced observables. 
This implies that different regions in  the SMC 
may exhibit statistical signatures of either 
compressive or solenoidally driven turbulence.
%, with compressive forcing mostly 
%occurring in swept-up shell regions.

\subsection{Summary}

\label{sec:conclusions}

We have investigated a new method for constraining turbulent properties of the ISM,
specifically the sonic Mach number, by using the HI column density image and a database of
numerical simulations with a range of sonic and Alf\'venic Mach numbers.
By applying the 3rd and 4th statistical moments on both observed and simulated data
we have derived the spatial distribution of the sonic Mach number across the 
SMC with angular resolution of 30$'$.
To provide an estimate of the Alf\'venic Mach number we used two approaches: 
the spatial power spectrum and the bispectrum.
Using the database of numerical simulations we have shown that the spatial power spectrum
varies with both the sonic and Alf\'venic Mach numbers. If the sonic number is known
the Alf\'venic number can be constrained from this dependence.
The bispectrum shows the level of correlation between turbulent eddies of different size and
depends greatly on the sonic Mach number, and somewhat on the Alf\'venic Mach number. By
comparing the bispectra of observations and simulations we have gauged the importance of
magnetic fields relative to the gas pressure in the SMC.
The following results were discussed in the paper.

\begin{itemize}
\item Skewness and kurtosis of the HI column density generally correlate well and
are within the range expected from MHD simulations. This suggests that 
departures from Gaussianity could be interpreted as being governed by
MHD turbulence. 

\item Most of the HI in the  SMC bar and the Eastern Wing is subsonic or
transonic with ${\cal M}_{s}\sim0-2$. Sites of most recent
star formation have ${\cal M}_{s}\sim1$. Regions with the highest skewness and kurtosis,
which could be interpreted as having ${\cal M}_{s}\sim4$, correspond to the edges of the bar.
The most turbulent regions are most likely tracing tidal or shearing flows. 
The fraction of the SMC with different turbulent properties is:
10\% with ${\cal M}_{s}>2$, 80\% with $0<{\cal M}_{s}<2$, and about 10\%
with very low values of ${\cal M}_{s}$.

\item Using HI absorption profiles from Dickey et al. (2000) we have estimated
that the CNM in the SMC is highly supersonic with ${\cal M}_{s}=3.5-4$. This is at least
a factor of two higher than what we measured from the higher statistical moments for 
the HI gas seen in emission. One possible reason for this discrepancy could be that
HI emission is dominated by warm gas and the fraction of the CNM in the SMC is $\la20$\%.

\item The slope of the spatial power spectrum and the bispectrum suggest that the HI in the
SMC is super-Alf\'venic with ${\cal M}_{A}\sim1-3$. This is implies that the gas
pressure dominates over the magnetic pressure.

\item The bispectrum of the HI column density shows large scale wave correlations suggesting a 
large scale energy injection mechanism. Contrary to simulations which show a smooth
decrease of wave-wave correlations from large to small scales, the SMC bispectrum
shows localized enhancements of correlations and at least one prominent break
at $\sim160$ pc. We speculate that the multi-phase medium, and/or energy injection by processes
other than turbulence, could be responsible for the correlation jumps. The break on the other
hand appears at a scale similar to the diameter of the majority of expanding shells in the SMC.

\end{itemize}

\acknowledgments
B.B is thankful for valuable discussions with Diego Falceta-Gon\c calves and Jungyeon Cho. 
BB acknowledges the NASA Wisconsin Space Grant Consortium and the National Science 
Foundation Graduate Research Fellowship. 
AL acknowledges NSF grant AST 0808118 and the Center for Magnetic Self-Organization.
SS acknowledges support from the NSF grant AST 0707679 and the Research Corporation.

\appendix
Here we present a technique that could be used to further illuminate subsonic regions in the ISM.

\begin{figure*}[tbh]
\centering
\includegraphics[scale=.6]{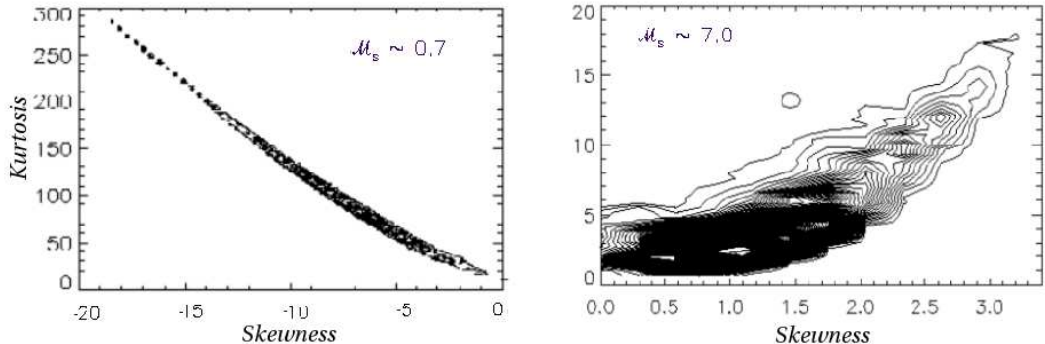}
\caption{Kurtosis vs. Skewness for isothermal simulations.  The left is subsonic 
while the right is supersonic. These figure are similar to~\ref{evidence} except that the beam was modified 
with 7 extra points set to zero.  
The supersonic models remain relatively unchanged in their trend while subsonic models with 
the modified distribution obtain a very tight anti-correlation due to extra points pushing them 
away from Gaussianity.  The supersonic model is unaffected because it is already very 
positively skewed. This could be employed as an additional method to detect subsonic 
gas using moments}
\label{evidence2}
\end{figure*}

\begin{figure*}[tbh]
\centering
\includegraphics[scale=.37]{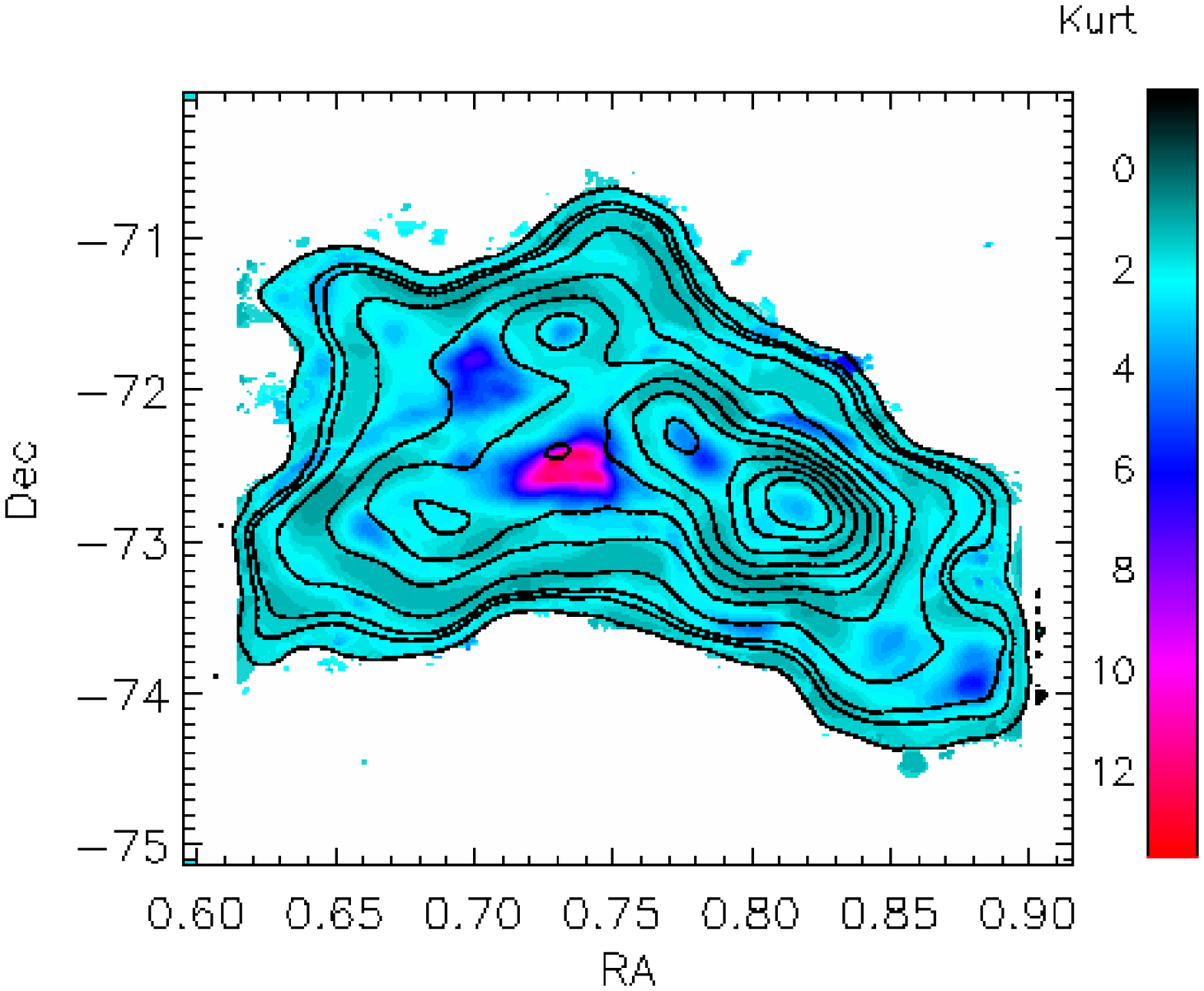}
\includegraphics[scale=.37]{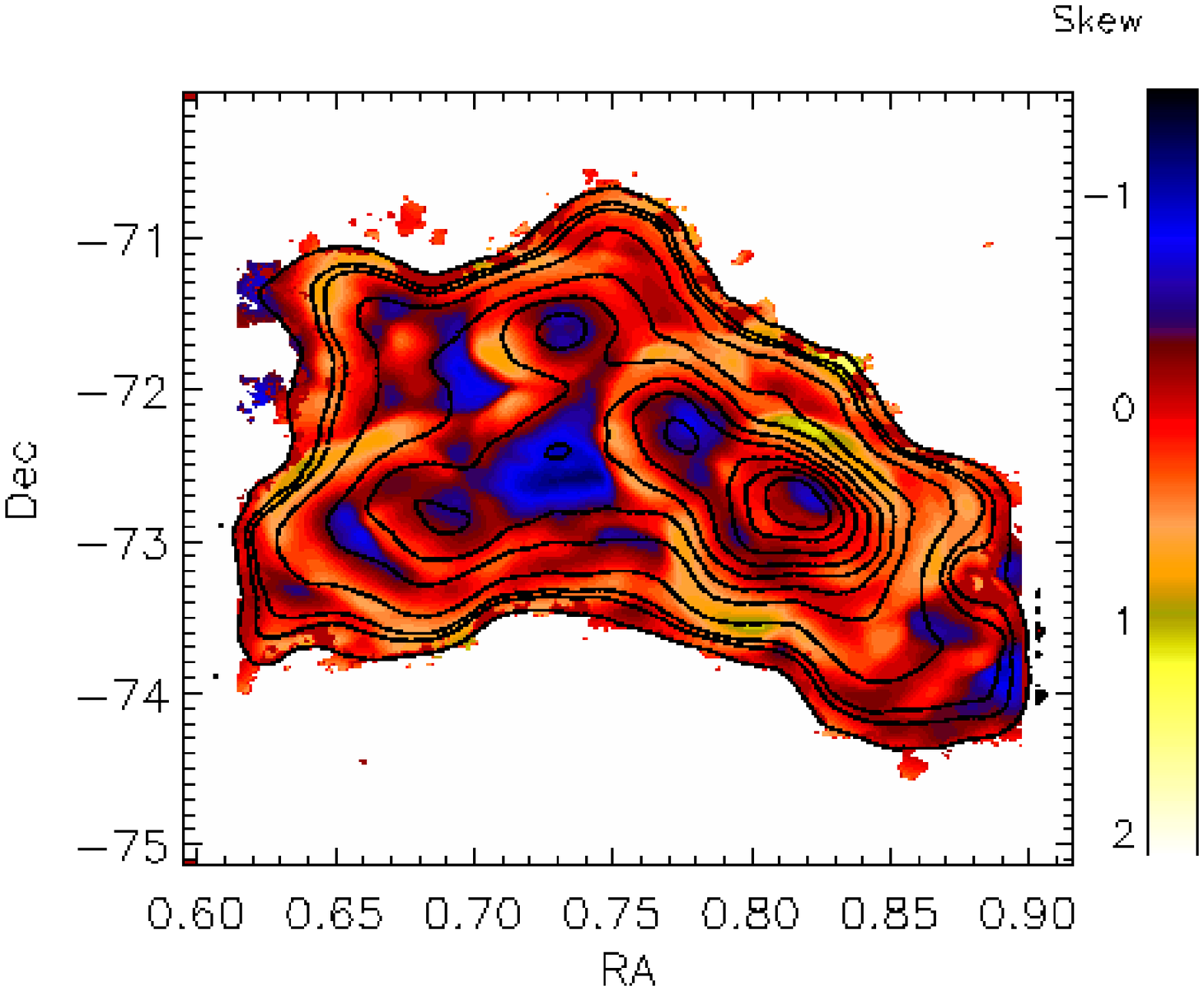}
\caption{Kurtosis and Skewness map with beams modified with 7 zero points side by side.  
Areas where they do not agree in sign (dark blue in skewness and dark blue/red in kurtosis) 
point to anti-correlation , which indicates regions that are potentially subsonic. 
These regions correspond to areas of
low kurtosis and skewness in the previous maps, yet are more clearly seen here. 
The largest of these subsonic regions is seen along the bridge
between the bar and the east wing (See Figure \ref{evidence}).}
\label{fig:compare}
\end{figure*}

In Figure~\ref{evidence2} we plot kurtosis vs. skewness in a manner similar to Figure~\ref{evidence}. 
The difference between this figure and Figure~\ref{evidence} is that we added seven zero 
points into the beam  (instead of 3841 points there are now 3848 points).  
This produces almost no change in the supersonic models since they already have high positive 
skewness and kurtosis. Supersonic models still show strong correlation between skewness and kurtosis.

 However a big change is seen in the subsonic skewness vs. kurtosis.  
The additional zero points shift the distribution from Gaussian (the mean of our simulations with no scaling applied is unity) 
to negative skewness and very peaked kurtosis producing a tight anti-correlation. 
Because this technique only strongly affects subsonic areas we can use it to locate subsonic turbulence in the 
SMC with skewness and kurtosis by looking for anti-correlation and very high values. 
 These properties also hold 
for simulations with cloud boundaries imposed. Note that one must use caution here and 
carefully examine the distribution 
of data.

The application of this technique to the SMC data is shown in 
Figure~\ref{fig:compare}.  This plot shows the skewness and kurtosis maps
of the SMC side by side with the modified beam.  Indeed, most regions are unchanged from the analysis in Section~\ref{sec:mmaps}. However 
a few regions stick out with signatures that are subsonic, that is, anti-correlation between skewness and kurtosis.  Two regions of the highest kurtosis are located in the area between the HI bar and the
Eastern Wing.
While kurtosis reaches values of 4-8, the corresponding skewness values are negative,
$\sim-1$. Again, such combination of skewness and kurtosis may correspond to simulations
of subsonic isothermal MHD turbulence, and/or points out additional processes at work.


\begin{thebibliography}{}
 %\bibitem[Barnett(2002)]{Bar02}
  %Barnett, A. G. 2002, PhD Thesis, School of Physical Sciences, The University of Queensland.
 \bibitem[Beresnyak, Lazarian \& Cho(2005)]{beresnyak05}
  Beresnyak, A., Lazarian, A. \& Cho, J., 2005, \apj, 624, L93
 \bibitem[Biskamp(2003)]{bis}
 Biskamp, D. (2003), Magnetohydrodynamical Turbulence, (Cambridge University Press, Cambridge). 
 \bibitem[Brunt \& Heyer(2002)]{brunt02}
  Brunt, C., \& Heyer, M.,2002,ApJ, 566, 27 
  \bibitem[Burkert(2006)]{bur06}
  Burkert, A., 2006, C. R. Physique 7 
 \bibitem[Burkhart et al.(2009)]{Burkhart08}
  Burkhart, B. Falceta-Goncalves, D., Kowal, G., Lazarian, A., 2009, \apj, 693, 250 (BFKL)
 \bibitem[Cho \& Lazarian(2002)]{Cho02}
  Cho, J. \& Lazarian, A. 2002, Phys. Rev. Lett., 88, 5001
 \bibitem[Cho \& Lazarian(2003)]{Cho03}
  Cho, J. \& Lazarian, A., 2003, \mnras, 345, 325
 \bibitem[Chepurnov \& Lazarian(2009)]{Chep09}
 Chepurnov, A., \& Lazarian, A., 2009, ApJ, 693, 1074
 \bibitem[Chepurnov et al.(2008)]{Chep08}
  Chepurnov, A., Lazarian, A., Gordon, J., \& Stanimirovic., S., ApJ, 688, 1021
\bibitem[Crovisier \& Dickey(1983)]{Crov83}
 Croviser, J., \& Dickey, M., 1983, A\&A, 122, 282
 \bibitem[Deshpande(2000)]{Deshpande00}
  Deshpande et al.,2000, ApJ, 543, 227
 \bibitem[Dickey et al.(2000)]{Dick00}
  Dickey et al., 2000, ApJ, 536, 756
 %\bibitem[Elmegreen(1999)]{Elm99}
  %Elmegreen, B. G. 1999, ApJ, 527, 266
 %\bibitem[Elmegreen \& Falgarone(1996)]{Elm96} 
 %Elmegreen, B.G., \& Falgarone, E.\ 1996, \apj, 471, 816
 \bibitem[Elmegreen \& Scalo(2004)]{Elm04}
  Elmegreen, B., \& Scalo, J., ARA\&A, 42, 211
 \bibitem[Esquivel \& Lazarian(2005)]{Esq05}
 Esquivel, A., \& Lazarian, A., 2005, ApJ, 631, 320
\bibitem[Esquivel \& Lazarian(2009)]{Esq09}
  Esquivel, A., \& Lazarian, A., 2009, ApJ in press
 \bibitem[Esquivel et al.(2003)]{Esq03}
  Esquivel, A., Lazarian, A., Pogosyan, D., \& Cho, J., 2003, MNRAS, 342, 325 
 \bibitem[Esquivel et al. (2007)]{}
  Esquivel et al., 2007, MNRAS, 381, 1733
 \bibitem[Federrath et al.(2009)]{Fed09}
  Federrath, C., et al.,2009, AA, in press
 \bibitem[Fry(1998)]{Fry98}
  Fry, J. N. 1998, Annals of the New York Academy of Sciences 848, 62
 \bibitem[Gill \& Henriksen(1990)]{Gill90} 
 Gill, A.G., \& Henriksen, R.N., 1990, \apjl, 365, L27
\bibitem[Goodman et al.(2009)]{Goodman09}
 Goodman, A. et al.,2009, ApJ, 692, 91
 \bibitem[Hatzidimitriou(2005)]{Hatz05}
 Hatzidimitriou et al., 2005, MNRAS, 360,1771
 \bibitem[Heiles \& Troland(2003)]{Heiles03}
 Heiles, C., Troland, T., 2003, ApJ,  586, 1067
 \bibitem[Heyer \& Zwiebel(2004)]{heyer04}
  Heyer, M., \& Zwiebel, E., 2004, Ap\&SS, 292, 9
 \bibitem[Hill et al.(2008)]{Hill08}
  Hill, A. S., Benjamin, R. A., Kowal, G., Reynolds, R. J., Haffner,. L. M. \& Lazarian, A., 2008, ApJS
 %\bibitem[Hinich \& Wilson(1990)]{Hinch90}
  %Hinich, M., \& Wilson, G., 1990, IEEE, 38, 1126
 \bibitem[Intrator et al.(1989)]{Intrator89} 
  Intrator, T., Meassick, S., Browning, J., Majeski, R. \& Hershkowitz, N. 1989, Phys. Fluids B, 1, 271-273
 \bibitem[Kennicutt(1995)]{Kennicut95}
  Kennicutt et al. 1995, AJ, 109, 594
 \bibitem[Kowal, Lazarian \& Beresnyak(2007)]{Kowal07}
  Kowal, G., Lazarian, A. \& Beresnyak, A., 2007, \apj, 658, 423  (KLB)
 \bibitem[Kowal et al.(2009)]{Kowal09}
  Kowal et al., 2009, ApJ, 700,63
 \bibitem[Kritsuk(2007)]{krit07}
 Kritsuk et al., 2007, ApJ, 665, 416
  \bibitem[Lazarian(2007)]{Laz07}
  Lazarian, A., 2007, JQSRT, 106, 225
 \bibitem[Lazarian(2009)]{Lazarian09} 
  Lazarian, A., 2009, SSR, 143, 357
  \bibitem[Lazarian et al.(2001)]{Laz01}
  Lazarian et al.,2001, ApJ, 555, 130
 \bibitem[Lazarian \& Pogosyan(2004)]{Lazarian04}
  Lazarian, A. \& Pogosyan, D., 2004, \apj, 616, 943
 \bibitem[Lazarian \& Pogosyan(2006)]{Lazarian06}
  Lazarian,A. \& Pogosyan,D., 2006,\apj, 652, 1348
 \bibitem[Lazarian \& Pogosyan(2000)]{Lazarian00}
  Lazarian, A. \& Pogosyan,D., 2000, \apj, 537, 720
 \bibitem[Lazarian \& Vishniac(1999)]{Laz99}
  Lazarian, A., \& Vishniac, E., 1999, ApJ, 511, 193
 \bibitem[Levy, Puppo \& Russo(1999)]{levy99}
  Levy, D., Puppo, G. \& Russo, G., 1999, Mathematical Modeling and Numerical Analysis, 33, 547
 \bibitem[Liguori et al.(2006)]{Liguori06} 
  Liguori, M., Hansen, F. K., Komatsu, E., Matarrese, S. \& Riotto, A., 2006, PhRvD, 73, 3505
 \bibitem[Lithwick \& Goldreich(2001)]{lithwick01}
  Lithwick, Y. \& Goldreich, P., 2001, \apj, 562, 279
 \bibitem[Liu \& Osher(1998)]{liu98}
  Liu, X.-D. \& Osher, S., 1998, Journal of Computational Physics, 141, 1
 %\bibitem[Mac Low \& Klessen(2004)]{mac04}
  %Mac Low, M.-M. \& Klessen, R. S. 2004, Rev. of Mod. Phys., 76, 125
% \bibitem[Masahiro \& Bhuvnesh(2004)]{Masa04}
 % Masahiro, T. \& Bhuvnesh, J. 2004, MNRAS, 348, 897
 \bibitem[Mao et al.(2008)]{Mao08}
  Mao et al. 2009, ApJ, 688, 1029
 \bibitem[McCray \& Snow(1979)]{mcc1979}
 McCray, R.,\& Snow, P, 1979, A\&AA, 17, 213
 \bibitem[McKee \& Ostriker(2007)]{mckee07} 
  McKee, C., Ostriker, E., 2007, 
%McKee, C. F. \& Tan, J. C., 2002, Nature, 416, 59
\bibitem[Monin, A.S.\& Yaglom, A.M.(1967)]{mon}
  Monin, A.S.\& Yaglom, A.M.,1967, Statistical Fluid Mechanics, MIT Press
 \bibitem[Montgomery et al.(1987)]{Mont87}
  Montgomery, D., Brown, M. R., Matthaeus, W. H., 1987, J. Geophys. Res., 92, 282
\bibitem[Nordlund \& Padoan(1999)]{Nor99}
  Nordlund, A. K. \& Padoan, P., 1999, Interstellar Turbulence, proceedings of the 2nd Guillermo Haro Conference, Cambridge University Press
\bibitem[Ostriker et al.(2001)]{Ostriker01}
  Ostriker, E. C., Stone, J. M. \& Gammie, C. F., 2001, ApJ, 546, 980
 %\bibitem[Padoan et al.(2006)]{Pad06}
 % Padoan, P, Juvla, M., Kritsuk, A., \& Norman, M.,2006, ApJ, 653, 125 
 %\bibitem[Padoan et al.(2003)]{Padoan03}
  %Padoan, P., Boldyrev, S., Langer, W., \& Nordlund, A., 2003, ApJ, 583, 308
 %\bibitem[Padoan et al.(2003)]{pa03} 
  %Padoan, P., Goodman, A., \& Juvela, M., 2003, ApJ, 588, 881
 %\bibitem[Putman et al.(1998)]{Putman98}
 % Putman et al., 1998, Nature, 394, 752
 %\bibitem[Scalo(1987)]{Scalo87}
  %Scalo, J. M. 1987, Interstellar Processes, ed. D. J. Hollenbach \& H. A. Thonson (Dordrecht: Reidel), 347
 \bibitem[Scoccimarro(2000)]{Scoccimarro00}
  Scoccimarro, R., 2000, \apj, 544, 597
% \bibitem[Shay et al.(2008)]{Shay08}
 % Shay et al., 2008, AGU, SM31A-1709
 \bibitem[Stanimirovic et al.(1999)]{SX99}
  Stanimirovic, S., Staveley-Smith, L., Dickey, J. M., Sault, R. J., \& Snowden,S. L., 1999, MNRAS, 302, 417 (SX99)
 \bibitem[Stanimirovic \& Lazarian(2001)]{Stan01}
  Stanimirovic, S. \& Lazarian, A., 2001, ApJ, 551, L53
 \bibitem[Stanimirovic(2001)]{Stan201}
  Stanimirovic, S., 2001, \textit{Astrophysics and Space Science}, 277, 87
 \bibitem[Stavely-Smith et al.(1997)]{Stavley97}
  Staveley-Smith, L., Sault, R. J., Hatzidimitriou, D., Kesteven, M. J., \& McConnell, D. 1997, MNRAS, 289, 225
 \bibitem[Stanimirovi{\'c}\& Staveley-Smith(2004)]{Stanimirovic04}
  Stanimirović, S., Staveley-Smith, L., \& Jones, P. A. 2004, ApJ, 604, 176
 \bibitem[Stanimirovi{\'c}(2007)]{stan07} 
  Stanimirovi{\'c}, S.,2007, IAU Symposium, 237, 84 
 %\bibitem[Stutzki(1994)]{stutzki94} 
  %Stutzki, J., 1994, Infrared Physics and Technology, 35, 493
 %\bibitem[Stutzki et al.(1998)]{stutzki98} 
  Stutzki, J., Bensch, F., Heithausen, A., Ossenkopf, V., \& Zielinsky, M., 1998, \aap, 336, 697
 \bibitem[Tabachnick \& Fidell(1996)]{Taba96}
  Tabachnick, B. G., \& Fidell, L. S. (1996). Using multivariate statistics (3rd ed.). New York: Harper Collins.
 \bibitem[Tynan et al.(2001)]{Tynan01} 
  Tynan, G. R., Moyer, R. A., Burlin, J. \& Holland, C. 2001, Phys. Plasmas, 8, 2691
 \bibitem[Westerlund(1991)]{West91}
Westerlund, B. E. 1991, in The Magellanic Clouds, IAU Symp., 148, 15 

\end{thebibliography}
\end{document}